\documentclass[         %
aps,                    
prd,                    
showpacs,               
nofootinbib,            
showkeys,               %
preprintnumbers,        %
floatfix]               
{revtex4}               
\usepackage{graphicx,longtable}

\begin{document}
\title{Supernova Relic Electron Neutrinos and anti-Neutrinos in future Large-scale Observatories}
\author{C. Volpe}
\email{volpe@ipno.in2p3.fr}
\author{J. Welzel}
\email{   welzel@ipno.in2p3.fr}
\affiliation{Institut de Physique Nucl\'eaire, F-91406 Orsay cedex,
France}

\date{\today}
\begin{abstract}
We investigate the signal from supernova relic neutrinos
in future large scale observatories, such as MEMPHYS 
(UNO, Hyper-K), LENA and GLACIER, at present under study.
We discuss that complementary information might be gained
from the observation of
supernova relic electron anti-neutrinos and neutrinos using the scattering on protons on one hand, and on nuclei such as oxygen,
carbon or argon on the other hand. When determining the relic neutrino fluxes
we also include, for the first time, the coupling of the neutrino magnetic moment to magnetic fields 
 within the core-collapse 
supernova. We present numerical results on 
both the relic $\nu_e$ and $\bar{\nu}_e$  fluxes and on the number of events for 
$\nu_e + ^{12}$C, $\nu_e + ^{16}$O, $\nu_e + ^{40}$Ar and $\bar{\nu}_e +$p for various oscillation scenarios.  
The observation of supernova relic neutrinos might provide us with unique 
information on core-collapse supernova explosions, on the
star formation history and on neutrino properties, that still remain unknown.
\end{abstract}
\medskip
\pacs{}
\keywords{core-collapse 
supernovae, supernova relic neutrinos, neutrino magnetic moment,  resonant spin-flavour precession}
\preprint{}
\maketitle

\section{Introduction}
\noindent
The observation of neutrinos from past core-collapse supernovae would represent an exciting and important 
discovery. Supernova relic neutrinos (SRN) bring crucial information on the explosion phenomenon, 
on the star formation rate and on neutrino properties. The understanding of how massive stars explode constitutes one
of the main future challenges in astrophysics. While the various stages of the stellar evolution are
quite well established, it still remains a mistery how, in these stars, 
the shock wave formed in the iron core, after collapse, manages
to eject the mantle. Essentially all the gravitational energy is emitted in a short burst of neutrinos of all flavours.
This luminosity curve follows closely the collapse, accretion phase and the cooling of the proto-neutron star left \cite{Raffelt:wa}. 

The first time such neutrinos have been observed is from the SN1987A located in the Large Magellanic Cloud,
thanks to the IMB \cite{Bionta87}, Baksan \cite{Alexeyev87} and Kamiokande detectors \cite{Hirata87}. 
The few electron anti-neutrino events measured have roughly confirmed the theoretical expectations,
even though their energy and angular distributions still have unsolved problems.  
For the SRN fluxes, the best upper limits at 90 $\%$ C.L. of 6.8 $\times$ 10$^3$ $\nu_e$ 
cm$^{-2}$s$^{-1}$  (25 MeV $< E_{\nu_e} <50$ MeV) and 1.2 $\bar{\nu}_{e}$ cm$^{-2}$s$^{-1}$  
($E_{\bar{\nu}_e} > 19.3$ MeV) come from the LSD \cite{Aglietta:1992yk} and the Super-Kamiokande detectors
\cite{Malek:2002ns}. The latter appears extremely close to most of theoretical predictions. Indeed it seems quite likely that the supernova relic neutrino background could be observed either by using an improved technology, e.g. the use of Gadolinium doped water \v{C}erenkov 
detectors \cite{Beacom:2003nk}, or 
by scaling the present technology at megaton sizes, like in MEMPHYS \cite{deBellefon:2006vq}, UNO \cite{uno}, and Hyper-K \cite{itow}. These large-size observatories could provide us with the complementary information from a possible
future (extra-)galactic explosion and from the SRN background.

Many authors have estimated the expected SRN flux using different approaches to determine the supernova rate
\cite{Malaney:1996ar,Hartmann:1997qe,Totani:1995dw,Kaplinghat:1999xi,Strigari:2003ig,Fukugita:2002qw,Wurm:2007cy} 
(see \cite{Ando:2004hc} for a review). In fact, the latter can be deduced either from direct observations \cite{Cappellaro:1996cc} or from star formation rates (see e.g. \cite{Strigari:2005hu}). 
The effects of neutrino masses and mixings on the SRN flux
have been included recently (see e.g. \cite{Ando:2002ky}), \cite{Ando:2002zj}). 
In \cite{Lunardini:2005jf} a pragmatic approach has been used: to predict the SRN flux using available data, namely direct supernova rate measurements and supernova neutrino fluxes
that give the best fit of the SN1987A events.
Most of the available literature presents calculations for the $\bar{\nu}_e$ scattering on protons, which is the dominant detection channel in water \v{C}erenkov (see e.g. \cite{Ando:2004sb}) or scintillator detectors \cite{Wurm:2007cy}. 
In particular, a study of what could be learned on the neutrino spectra and/or the core-collapse supernova rate at
next generation water \v{C}erenkov detectors is realized in \cite{Lunardini:2006pd}.
In Ref. \cite{Cocco:2004ac} 
the potential of liquid argon detectors
to observe supernova relic electron neutrinos is analyzed, while
the specific case of $\nu_e$ detection in SNO is considered in \cite{Beacom:2005it,Lunardini:2006sn}.
Ref. \cite{Lunardini:2006sn} also obtains an upper limit on the relic $\nu_e$ flux from the $\bar{\nu}_e$ upper limit in Super-Kamiokande. 
Finally, in 
\cite{Ando:2003ie,Fogli:2004gy} the authors have 
considered the possibility of using the SRN to gather information on more exotic unknown neutrino properties such as neutrino decay.

The study presented in this paper has several new features compared to the available literature. First we study the possible signal from both the supernova relic $\nu_e$ and $\bar{\nu}_e$ 
in future large-scale observatories.
We consider the three technologies under investigation, water \v{C}erenkov, scintillators and liquid argon, having in mind the proposed MEMPHYS  (UNO, Hyper-K), LENA and GLACIER detectors \cite{Autiero:2007zj}. 
We also include for the first time in the neutrino propagation within the supernova the interaction
with matter (the Mikheev-Smirnov-Wolfenstein effect \cite{Wolfenstein:1977ue,Mikheev:1986wj}) 
as well as the coupling of the neutrino magnetic moment to the magnetic fields (the spin-flavour conversion phenomenon). 
Clearly our present understanding of supernovae explosion is still limited from several respects
and, in particular, for the knowledge of the stellar magnetic fields and their evolution during the explosion. We think however that, even at the present stage, it is interesting 
to perform a first calculation including this process, to start investigating under which conditions some effects might be seen. We make here the assumption of a static supernova density profile,
neglecting in this first work the changes induced by the shock wave. We also do not consider earth matter effects that have been discussed in \cite{Ando:2002zj}. 
To determine the SRN flux, we integrate the neutrino
spectra from one supernova explosion over the redshift by
convoluting it with the supernova explosion rate. 
We predict the expected event rates and
analyse the effects of the presence of a neutrino magnetic moment both on the relic number fluxes
and on the number of events. Note that  we follow previous investigations for
the choice of the neutrino fluxes at the neutrinosphere, of the supernova rate as well as 
for the discussion of the possible backgrounds in the detectors.
The paper is organized as follows. The formalism is described in Section II.
The results and conclusions are presented in Sections III and IV.

\section{Formalism}
\subsection{Neutrino evolution in a core-collapse supernova with resonant spin-flavour conversion}
\noindent
Neutrino evolution including coupling with matter and with stellar magnetic fields  
 (the resonant spin-flavour conversion or RSF) was first considered as a possible solution of the solar neutrino
deficit problem \cite{Cisneros:1970nq,Lim:1987tk,Akhmedov:1988uk}. 
We now know that this deficit is due to the presence of neutrino masses and
mixings \cite{Yao:2006px}. Even though the RSF might still give a sub-dominant contribution, it turns out to be extremely small \cite{Balantekin:2004tk}.
The situation could be quite different in the case of massive stars such as core-collapse supernovae (SNII), 
since strong magnetic fields are expected, particularly close to the proto-neutron star surface. 
Many authors have investigated the RSF effects 
in SNII (see e.g. 
\cite{Lattimer:1988mf,Nunokawa:1996gp,Ando:2002sk,Akhmedov:2003fu,Balantekin:2007xq}).

The knowledge of the neutrino magnetic moment offers a serious challenge\footnote{Note that since neutrinos are massive
 particles they certainly have a non-zero magnetic moment. }. While the minimal 
extension of the Standard Model predicts a very small value \cite{Raffelt:wa}, the observation of a
large value would clearly indicate new physics. Several bounds have been obtained so far.  
Direct upper limits come from neutrino-electron scattering using neutrinos
from reactors \cite{Reines:1976pv,texono,Daraktchieva:2005kn,Yao:2006px}, 
the best value being $9. \times 10^{-11}~\mu_B$  at $90 \%$ C.L. \cite{Daraktchieva:2005kn}.
Based on the electroweak radiative corrections to the neutrino mass, a model-independent
theoretical upper bound is derived in \cite{Bell:2006wi}. 
Important indirect limits are obtained from astrophysical observations such as
red-giant cooling \cite{Raffelt:1990pj} 
the SN1987A \cite{Lattimer:1988mf}, or Big-Bang nucleosynthesis \cite{Morgan:1981zy}. 
One might hope to attain improved direct limits
with running experiments at reactors, a very intense tritium source, 
low energy beta-beams \cite{McLaughlin:2003yg}
or future astrophysical observations.

The neutrino evolution equations for three flavours including the interaction with matter and the coupling to 
the magnetic fields is given by ($\hbar=c=1$):
\begin{equation}\label{e:1}
i \frac{d}{dr} \left(
	      \begin{array}{c} 
	       \nu \\
	       \bar{\nu}
	      \end{array}
	      \right)
= \left( \begin{array}{cc}
 H_0 & B M \\ -B M & \bar{H_0} \end{array} \right)
\left( \begin{array}{c} \nu \\ \bar{\nu} \end{array} \right),
\end{equation}
where $\nu$ ($\bar{\nu}$) stand for the electron, muon and tau (anti-)neutrinos.
The hamiltonian $H_0$ includes the neutrino mixing and interaction with matter:
 \begin{eqnarray}\label{e:2}
  H_0 = \frac{1}{2E_\nu} U \left(
			  \begin{array}{ccc}
			  0 & 0 & 0 \\
			  0 & \Delta m^2_{12} & 0 \\
			  0 & 0 & \Delta m^2_{13}
			  \end{array}
			  \right) U^\dagger
  + \left(
   \begin{array}{ccc}
   V_{cc}^e+V_{nc} & 0 & 0 \\
   0 & V_{cc}^{\mu}+V_{nc} & 0 \\
   0 & 0 & V_{cc}^{\tau}+V_{nc}  
   \end{array}
 \right), 
 \end{eqnarray}
\noindent
where the first term corresponds to neutrino propagation in vacuum while the second includes $V_{cc}^i$, the potential due to the charged-current  neutrino scattering with electrons and $V_{nc}$, the neutral-current neutrino scattering with nucleons. In Eq.(\ref{e:2}) 
$U$ stands for the Maki-Nakagawa-Sakata-Pontecorvo (MNSP) matrix relating the neutrino flavor 
and mass basis. 
For anti-neutrinos,  $\bar{H}_0$ is similar to $H_0$ but with a minus sign in front of  the second term. 
The off diagonal terms come from the neutrino interactions with the SNII magnetic field, $B(r)$ and $M$ is the magnetic moment matrix.
Here we consider the case of Majorana neutrinos for which M is  
 \begin{equation}
M=\left(
\begin{array}{ccc}\label{e:3}
   0 & \mu_{e\mu} &  \mu_{e\tau} \\
   - \mu_{e\mu} & 0 &  \mu_{\mu\tau} \\
   -\mu_{e\tau} &  -\mu_{\mu\tau} & 0  
   \end{array}
\right)
\end{equation}
with $\mu_{ij}$ the transition magnetic moments. (In \cite{Balantekin:2007xq} we have shown that 
neutrino flux modifications of less than 1 $\%$ are induced by the contribution of the neutrino magnetic moment to the neutrino scattering with matter. Here we do not consider these contributions.)
From the RSF point of view Eq.(\ref{e:1}) the relevant quantity is the product $B M$. 
The observation of RSF effects from a future SNII explosion signal, or
from supernova relic neutrinos, might provide us with further information on the value of $ \mu_{\nu}$ 
and/or the strength and profile of the magnetic fields in SNII.

Coupling with matter or with magnetic fields can induce important modifications to neutrino oscillations.
While the former
produce $\nu_e$ versus $\nu_{\mu}$ and $\nu_{\tau}$, the latter can produce
$\nu_e$ versus $\bar{\nu}_{\mu}$ and $\bar{\nu}_{\tau}$  resonant conversions. The MSW resonances are situated
at two densities governed by ($\Delta m^2_{13},\theta_{13}$) and   ($\Delta m^2_{12},\theta_{12}$) respectively. 
The first resonance is located at density of about $10^3$ g/cm$^3$ (MSW-$h$), while the second one at about 
1 g/cm$^3$ (MSW-$l$). The neutrino conversion at each resonance depends on its adiabaticity. While the 
MSW-$l$ resonance is adiabatic for typical supernova density profiles, the adiabaticity of MSW-$h$ depends
on the still unknown mixing angle $\theta_{13}$ \cite{Dighe:1999bi,Engel:2002hg}. Similarly, a resonant $\nu_i$ versus $\bar{\nu}_{j}$ conversion (and vice-versa)
can take place due to the presence of the neutrino magnetic moment. Three supplementary resonances are present in this case, 
their adiabaticy depending on the strength of the magnetic fields (see e.g. \cite{Akhmedov:2003fu} for a detailed discussion).

\subsection{Core-collapse supernova neutrinos and the supernova rate}
\noindent
Core-collapse supernovae are one of the most explosive phenomenon in our
Universe. When the iron core of a massive star reaches its Chandrasekhar limit, an increased
gravitational pull due to a contraction is no longer counterbalanced by the increase in
the thermal pressure. The iron core collapses
until it reaches nuclear densities and becomes very stiff. 
A shock wave is formed that moves outward
ejecting the stellar mantle. How such ejection takes place remains still an unknown issue.
Convection or extra energy deposit by neutrinos might offer the clue to this
longstanding problem. In fact, most of the energy produced by the explosion is 
taken away by a short burst of neutrinos of all flavours. Their luminosity curve follows closely
the different phases of the explosion. 

Two ingredients are crucial to predict the SRN differential flux, $d F/ d E_{\nu}$, i.e. the neutrino flux from one supernova and
the supernova rate : 
\begin{equation}\label{e:4}
\frac{d F}{d E_{\nu}}=\int^{z_{max}}_{0}\mathcal{R}_{SN}(z,\Omega_{\lambda},\Omega_m)\frac{d n_{\nu}((1+z) E_{\nu})}{d E_{\nu}}(1+z)\frac{d t}{d z}d z
\end{equation}
where $\mathcal{R}_{SN}(z,\Omega_{\lambda},\Omega_m)$ is the supernova rate per comoving volume at redshift $z$ ($\Omega_{\lambda}$ and $\Omega_m$ are cosmological parameters, $\Omega_{\lambda} + \Omega_m = 1$ in the standard $\Lambda$CDM model), 
$z_{max}$ is the redshift at which star formation started and $d n_{\nu}/d E_{\nu}((1+z) E_{\nu})$ is the energy spectrum of emitted neutrinos leaving the supernova \cite{Ando:2002ky}.

The neutrino spectra  $d n_{\nu}/d E_{\nu}$ at the neutrinosphere are quite well approximated by Fermi-Dirac
\cite{Raffelt:wa} or power-law distributions \cite{Keil:2002in}. Electron (anti)neutrinos interact with matter via charged as well as neutral current, 
while muon and tau neutrinos via neutral current only. The matter being neutron-rich and
the cross sections for anti-neutrinos weaker than for neutrinos, we expect the following neutrino hierarchy: 
$\langle  E_{\nu_e} \rangle < \langle  E_{\bar{\nu}_e} \rangle < \langle  E_{\nu_{x},\bar{\nu}_x} \rangle$ with
$x=\mu,\tau$. Supernova simulations predict typical temperatures of 10-13 MeV, 13-16 MeV and
16-23 MeV depending on the microphysics \cite{Keil:2002in}. 
To determine the $d n_{\nu}/d E_{\nu}$ at the SNII surface, 
we solve the neutrino evolution Eqs.(\ref{e:1}-\ref{e:3}), where
one of the important unknowns is the stellar magnetic field.
Here we consider that its shape follows a power-law,
as done e.g. in \cite{Nunokawa:1996gp}
\begin{equation}\label{e:5}
B(r)=B_0\left(\frac{r_B}{r}\right)^{n}
\end{equation}
where $n=2$ or 3 and $r_B=1.4 \times 10^{8}$cm, the iron-core surface.  

By including $dt/dz$ and using
\begin{equation}\label{e:6}
\mathcal{R}_{SN}(z,\Omega_{\lambda},\Omega_m)=(1+z)^{-3/2}\sqrt{(1+\Omega_m z)(1+z^2)-\Omega_{\lambda}(2z+z^2)}\mathcal{R}_{SN}(z)
\end{equation}
one gets that $d F/ d E_{\nu}$ is independent from the cosmological parameters:
\begin{equation}\label{e:7}
d F/ d E_{\nu} =\frac{1}{H_0}\int^{z_{max}}_{0}\mathcal{R}_{SN}(z)\frac{d n_{\nu}((1+z) E_{\nu})}{d E_{\nu}}(1+z)^{-3/2}d z
\end{equation}
where we take $z_{max}$ equal to 4.
For the supernova rate we consider the functional form from indirect observations accounting for the star formation rate
\cite{Ando:2002sk}
\begin{equation}\label{e:8}
\mathcal{R}_{SN}(z)=1.22\times10^{-2}\times 0.3 f_{\star}h_{65}\frac{\exp{(3.4z)}}{\exp{(3.8z)}+45}\ \mathrm{yr}^{-1}\mathrm{Mpc}^{-3}
\end{equation}
with $h_{65}=H_0/65 $km s$^{-1}$ Mpc $^{-1}$. The factor $1.22\times10^{-2}$ is the fraction of SNII, deduced by assuming the Salpeter IMF (with a lower cutoff at 0.5 $M_{\odot}$ and taking all stars with $M > 8 M_{\odot}$ as SNII).
The normalization factor $f_{\star}$ takes into account the uncertainty on the 
local star formation rate density, which at $z=0$ is (0.5-2.9)$\times 10^{-2} h_{70} M_{\odot}$ year$^{-1}$Mpc$^{-3}$. 
The value $f_{\star}=1$, that we will take in the following, is consistent with 
mildly dust-corrected UV data at low redshift \cite{Ando:2004hc} and predicts a local star formation rate of
0.7$\times 10^{-2} h_{70} M_{\odot}$ year$^{-1}$Mpc$^{-3}$. Note that according to \cite{Strigari:2005hu}, 
the local value for the star formation rate is  1.6$\times 10^{-2} h_{70} M_{\odot} $ year$^{-1}$Mpc$^{-3}$ which means $f_{\star}=2.5$.
Here we do not perform a study of the sensitivity of the SRN flux to different choices of the $\mathcal{R}_{SN}(z)$ since
this was performed in \cite{Ando:2002ky}. The different functionals available in the literature essentially have a very similar increasing behaviour for low redshifts
($z<1$), well constrained by observations; while they might differ at high redshifts ($z>1$). Note however that the latter contribute
little to the signal in an observatory, if the detection threshold is larger than about 10 MeV. 

\begin{figure}[t]
\vspace{.6cm}
\includegraphics[scale=0.3,angle=-90]{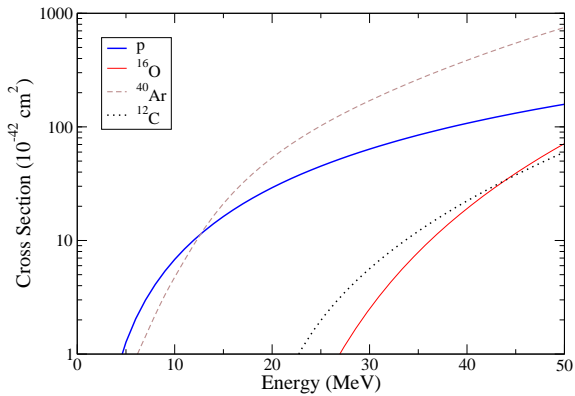}
\caption{Cross sections as a function of neutrino energy for $\bar{\nu}_e + p$ \cite{Balantekin:2006ga}
and $\nu_e + ^{12}$C \cite{Volpe:2000zn}, $\nu_e + ^{16}$O \cite{Lazauskas:2007bs} 
and $ \nu_e + ^{40}$Ar \cite{argon}. 
\label{fig:cross}}
\end{figure}

Once the relic fluxes are obtained, 
the event rates are calculated by multiplying the SRN flux by the corresponding cross sections 
(Figure \ref{fig:cross}):
\begin{equation}\label{e:9}
N_{events} = \int_{E_0} N_p dF/dE_{\nu}\sigma(E_{\nu}) dE_{\nu}
\end{equation}
with $E_0$ the threshold energy that depends on the detector technology, $N_p$ the number of target nucleons or nuclei, $\sigma$ the relevant cross section.

\section{Results}
\noindent
We have calculated the neutrino evolution in a core-collapse supernova Eqs.(\ref{e:1}-\ref{e:3}) and
then the relic supernova neutrino flux Eqs.(\ref{e:4}-\ref{e:8}). 
The oscillation parameters are fixed at 
$\Delta m^2_{12}= 8 \times 10^{-5}$eV$^2$, sin$^2 2\theta_{12}=0.86$ and
$\Delta m^2_{23}= 3 \times 10^{-3}$eV$^2$, sin$^2 2\theta_{23}=0.96$ for
the solar and atmospheric differences of the mass squares and
mixings, respectively \cite{Yao:2006px}. For the third still unknown neutrino mixing angle
$\theta_{13}$, we take either the present upper limit 
sin$^2 2\theta_{13}=0.19$ at 90 $\%$ C.L. (L) or  a small value of 
sin$^2 2\theta_{13}=10^{-4}$ (S) that might be attained at the future (third generation)  long-baseline
experiments \cite{Volpe:2006in}. 
Since the sign of  the
atmospheric mixing is unknown, we consider both the normal (NH) and
inverted (IH) hierarchy, i.e. $\Delta m_{13}^2 > 0$ or $<0$. 

\begin{figure}[t]
\vspace{.6cm}
\includegraphics[scale=0.3,angle=-90]{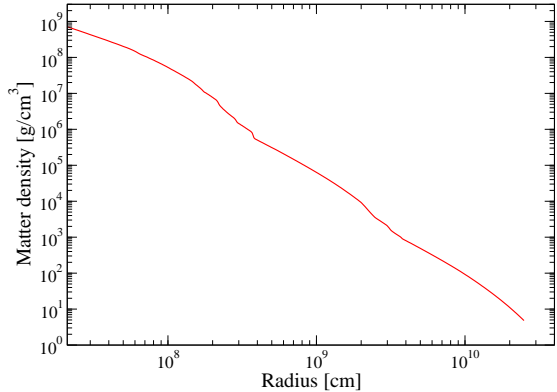}
\caption{Density profile of the pre-supernova progenitor used in this work \cite{Kneller:2007kg}. 
\label{fig:density}}
\end{figure}

For the neutrino spectra at the neutrinosphere,
we take Fermi-Dirac distributions as an example (Table \ref{tab:Tnu}).
We do not perform a detailed study of the sensitivity of the SRN flux predictions to different choices of the neutrino temperatures,
from available supernova simulations, since this was done in \cite{Ando:2004sb}.
The total luminosity is fixed at $L_0=3 \times 10^{53}$ ergs and
the energy is assumed to be equally shared among all flavours (this is not necessarily true at early times).
For the SNII matter density profile, we use a static 13.2 M$_{\odot}$ progenitor \cite{Kneller:2007kg} 
(Figure \ref{fig:density}). 
With the goal of also investigating the RSF effects on the SRN signal,
we consider two different power law profiles Eq.(\ref{e:5}) and 
the range of 0.001-1 values for the $\mu_{ij}B_0$ product Eqs.(\ref{e:1}-\ref{e:3}). These values correspond
to taking at $r=r_B$ for example a magnetic field of $B_0=10^{9}$ ($B_0=10^{12}$) Gauss and a transition magnetic moment of
$10^{-12}$ ($10^{-15}$) $\mu_B$ or $B_0=10^{12}$ Gauss and $10^{-12}$ $\mu_B$ respectively. In the following we will denote the largest value of 0.1 (1) for $n=2$ ($n=3$) 
by $(\mu B)_{large}$. Note that the radial profiles and the $\mu_{ij}B_0$ product are such that, 
while $\mu_{ij}B(r_B) =0.1,1$ for $n=2,3$; for both profiles $\mu_{ij}B(10 r_B)=0.001$ at 
the location of the RSF-$h$ resonance. However, the behaviour is different at the RSF-$l$ location. This has an impact on the expected rates, as we will see.

\begin{center}
\begin{table}[h!]
\begin{tabular}{|c|c|c|c|c|c|c|}
\hline
Electron fraction $Y_e$ & $T_{\nu_e}$ & 
$T_{\bar{\nu}_e}$ &$T_{\nu,\bar{\nu}_x}$ & $\left<E_{\nu_e}\right>$  & 
$\left<E_{\bar{\nu}_e}\right>$ & 
$\left<E_{\nu,\bar{\nu}_x}\right>$ \\ \hline
0.48 & 3.5 MeV& 5.1 MeV & 7.0 MeV & 11 MeV & 16 MeV & 22 MeV\\ \hline
\end{tabular}
\caption{Neutrino temperatures and average energies at the neutrinosphere.
\label{tab:Tnu}}
\end{table}
\end{center}

\subsection{The supernova relic $\nu_e$ and $\bar{\nu}_e$ fluxes}
\begin{figure}[t]
\vspace{.6cm}
\centerline{\includegraphics[scale=0.3,angle=-90]{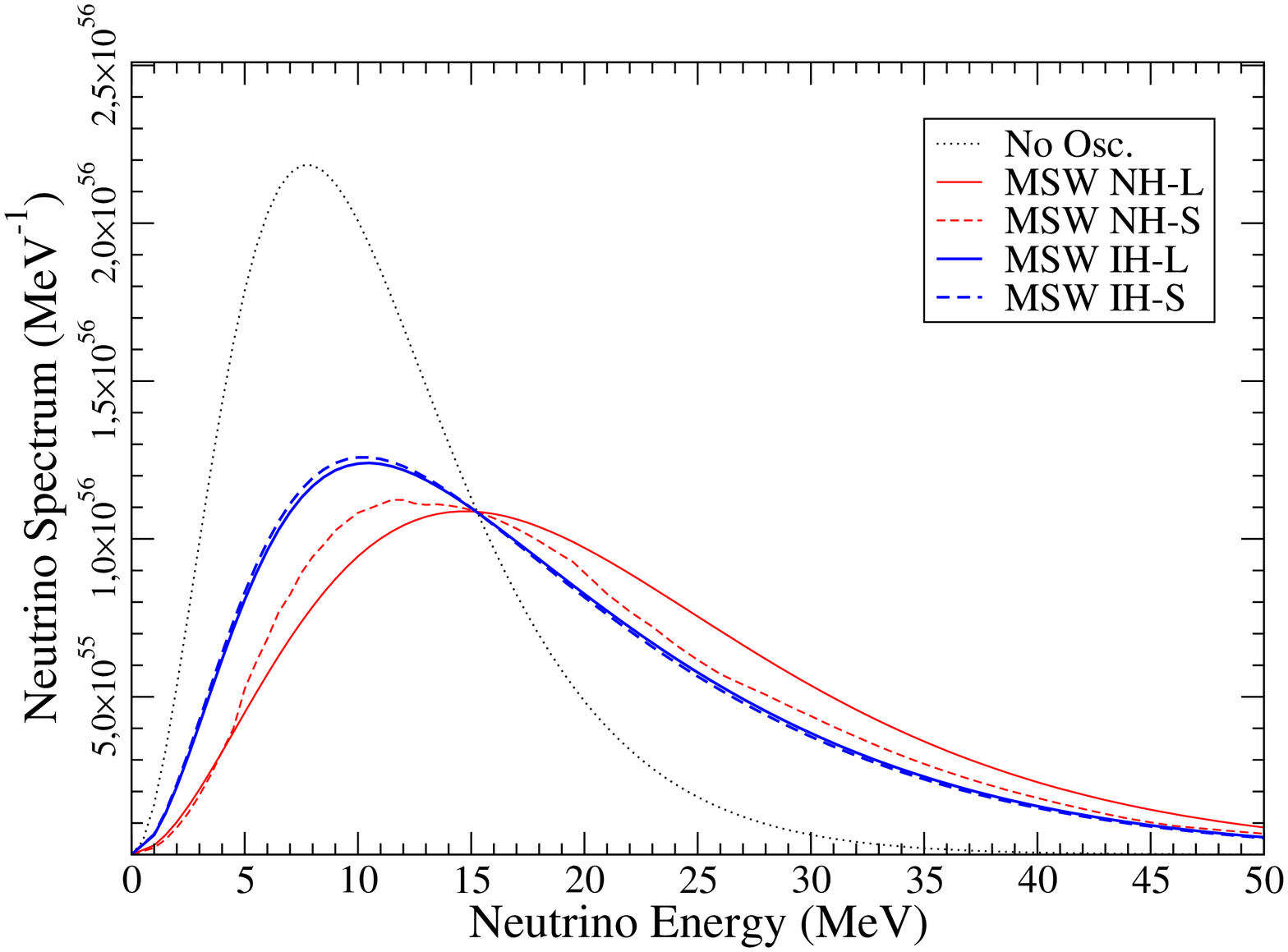}
\hspace{.2cm}\includegraphics[scale=0.3,angle=-90]{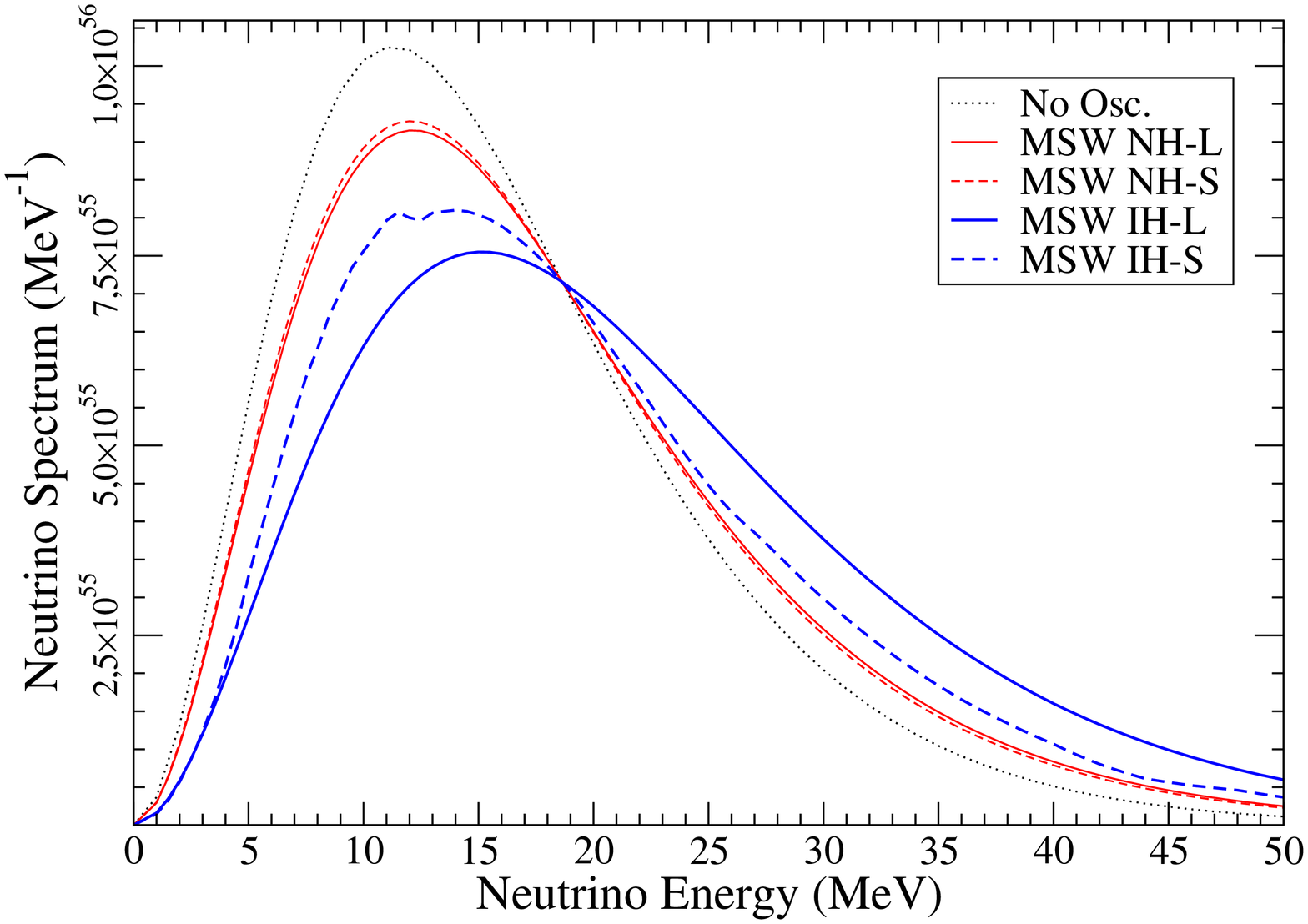}}
\caption{Electron neutrino (left) and anti-neutrino (right) spectra after propagation in a core-collapse supernova, 
Eqs. (\ref{e:1}-\ref{e:2}) with $B=0$, 
for the normal hierarchy with sin$^2 2\theta_{13}=0.19$ (NH-L) or sin$^2 2\theta_{13}=$10$^{-4}$ (NH-S)
(thin solid and dashed lines) and the inverted hierarchy with sin$^2 2\theta_{13}=0.19$ (IH-L) or 
sin$^2 2\theta_{13}=$10$^{-4}$ (IH-S) (thick solid and dashed lines). 
The Fermi-Dirac distribution is also shown for comparison (dotted line).
\label{fig:MSW}}
\end{figure}

\noindent
Figure \ref{fig:MSW} shows the neutrino spectra after propagation in 
the supernova for the two mass hierarchies and values of the third neutrino mixing angle 
(the magnetic fields or $\mu_{\nu}$ are set to zero). 
The corresponding average neutrino energies are shown in Table \ref{tab:aven}.
Note that the value of $\theta_{13}$ determines the
adiabaticity of the first MSW resonance at high density
\cite{Dighe:1999bi,Engel:2002hg}, while $\theta_{12}$ governs the
second (adiabatic) one at low density. As expected, when $\theta_{13}$ is large and the hierarchy is normal, 
maximal conversion occurs for
electron neutrinos which acquire the hot spectrum of muon and tau neutrinos. On the other hand, 
for inverted hierarchy (when there is no resonance), the electron neutrino spectra are much colder.
A similar behaviour is obtained for the electron anti-neutrinos, in the case of the inverted hierarchy. 
\begin{table}
\begin{center}
\noindent \begin{tabular}{|c|c|c|c|c|}
\hline
&NH-L&NH-S&IH-L&IH-S\\
\begin{tabular}{|c|}
\hline
 \\ \hline
MSW \\ \hline
$(\mu B)_{large}$ $n=2$\\ \hline
$(\mu B)_{large}$ $n=3$ \\ 
\end{tabular}
&
\begin{tabular}{|cc|}
\hline
$\left<E_{\nu_e}\right>$&$\left<E_{\bar{\nu}_e}\right>$\\ \hline\hline
21.5&17.7 \\ \hline
19.6&20.0\\ \hline
19.3&21.4\\
\end{tabular}
&
\begin{tabular}{|cc|}
\hline
$\left<E_{\nu_e}\right>$&$\left<E_{\bar{\nu}_e}\right>$\\ \hline\hline
18.8&17.4 \\ \hline
18.3&20.3\\ \hline
18.2&21\\
\end{tabular}
&
\begin{tabular}{|cc|}
\hline
$\left<E_{\nu_e}\right>$&$\left<E_{\bar{\nu}_e}\right>$\\ \hline\hline
17.9&21.8 \\ \hline
15.8&18.9\\ \hline
21.5&14.2\\
\end{tabular}
&
\begin{tabular}{|cc|}
\hline
$\left<E_{\nu_e}\right>$&$\left<E_{\bar{\nu}_e}\right>$\\ \hline\hline
17.6&18.5 \\ \hline
17.8&14.6\\ \hline
19&16\\
\end{tabular}
\\ \hline
\end{tabular}
\caption{Average energies (in MeV) of the $\nu_e$ and $\bar{\nu}_e$ fluxes, 
after the propagation in the SNII environment, with ($(\mu B)_{large}$ $n=2,3$) Eq.\ref{e:5} or without (MSW) the coupling to the 
magnetic fields. The different cases correspond to normal (NH) or inverted (IH) hierarchy and 
sin$^2 2\theta_{13}=0.19$ (L) or  
sin$^2 2\theta_{13}=10^{-4}$ (S). \label{tab:aven}}
\end{center}
\end{table}

\begin{figure}[h]
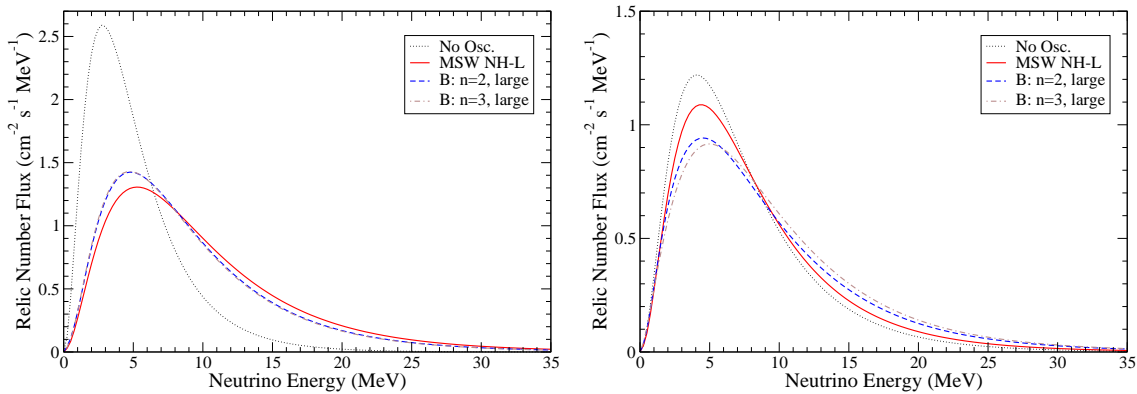

\vspace{.6cm}
\centerline{\includegraphics[scale=0.3,angle=0]{SRNue-NHL.eps}
\hspace{.2cm}\includegraphics[scale=0.3,angle=0]{SRNuebar-NHL.eps}}
\caption{Same as Fig. \ref{fig:MSW} but for the number flux of supernova relic neutrinos and anti-neutrinos
Eqs.(\ref{e:1}-\ref{e:8}) with and without the coupling to the magnetic fields. Their radial profile is taken as a power law Eq.(\ref{e:5}) with n=2,3 
and the product  $\mu_{ij}B$ is set to 1/0.1 for $n=3/2$ (see text). For values much smaller than 0.01 (0.1) for $n=2$ ($n=3$) the results are close to the MSW solution and they are not shown.  All curves correspond to
the NH-L case. Similar results are obtained for the NH-S.
\label{fig:RSN-NH}}
\end{figure}

\begin{figure}[t]
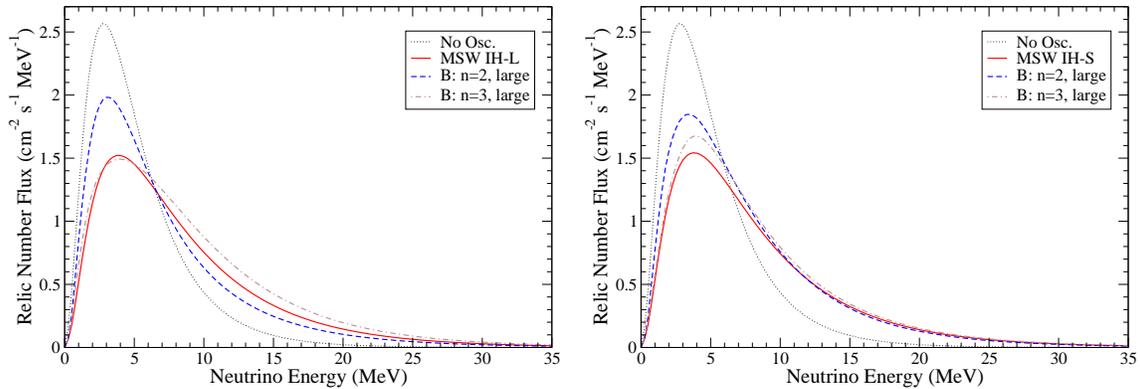

\vspace{.6cm}
\centerline{\includegraphics[scale=0.3,angle=0]{SRNue-IHL.eps}
\hspace{.2cm}\includegraphics[scale=0.3,angle=0]{SRNue-IHS.eps}}
\caption{Same as Fig. \ref{fig:RSN-NH} but for electron neutrinos in 
the IH-L (left) and IH-S (right) cases. 
\label{fig:RSNnue-IH}}
\end{figure}

\begin{figure}[t]
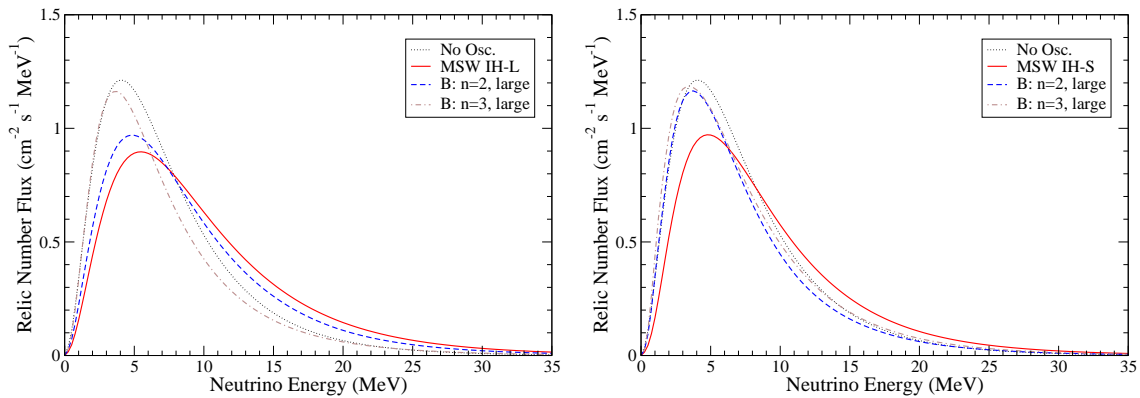

\vspace{.6cm}
\centerline{\includegraphics[scale=0.3,angle=0]{SRNuebar-IHL.eps}
\hspace{.2cm}\includegraphics[scale=0.3,angle=0]{SRNuebar-IHS.eps}}
\caption{Same as Fig. \ref{fig:RSNnue-IH} but for electron anti-neutrinos. 
\label{fig:RSNnuebar-IH}}
\end{figure}

The supernova relic $\nu_e$ and $\bar{\nu}_e$ fluxes, 
obtained after 
integrating over the redshift, are shown in Figures \ref{fig:RSN-NH} to \ref{fig:RSNnuebar-IH}. The results including
the RSF effects correspond to $\mu_{ij}B_0=0.1$ (1) for $n=2 $ ($n=3$). 
We do not show the supernova relic number fluxes obtained 
for values smaller than 0.001 (0.01) for $n=2$ ($n=3$) since the RSF effects are small and the results are close to the MSW solutions. For values within 0.001 (0.01) and 0.1 (1.) for $n=2$ ($n=3$) the results are in between the RSF and MSW ones. 
Table \ref{tab:rflux} presents the relic $\nu_e$ and $\bar{\nu}_e$ number fluxes integrated over different energy intervals.
One can see that the relic $\nu_e$ number flux is larger than the
$\bar{\nu}_e$ one in practically all cases\footnote{Note that our relic $\nu_e$ fluxes are compatible with the theoretical upper limit  of 5.5 cm$^{-2}$s$^{-1}$ extrapolated from the Super-Kamiokande result \cite{Lunardini:2006sn}.}. Since
the relic number fluxes are all peaked in the first 10 MeV, it is clear that this gives the most important contribution to the fluxes.
Therefore detector energy thresholds around 10 or 20 MeV -- as considered or used 
for example for LENA and water \v{C}erenkov detectors respectively -- are only sensitive to the tails of the SRN spectra.
In the energy interval larger than 19.3 MeV, our $\bar{\nu}_e$ relic flux ranges between 0.4-0.7 cm$^{-2}$s$^{-1}$, depending on the neutrino unknown parameters, and is compatible with 
the Super-Kamiokande
limit of 1.2 cm$^{-2}$s$^{-1}$ \cite{Malek:2002ns}. As far as the comparison to previous calculations is concerned, our
predictions agree well with e.g. \cite{Ando:2002ky},\cite{Strigari:2003ig}. 

\begin{table}
\begin{center}
\noindent \begin{tabular}{|c|c|c|c|c|}
\hline
&NH-L&NH-S&IH-L&IH-S\\
\begin{tabular}{|c|}
\hline
\\ \hline\hline
1 - 50 MeV \\ \hline
19.3 - 50 MeV \\ \hline
19.3 - 27 MeV\\ 
\end{tabular}
&
\begin{tabular}{|c|c|}
\hline
$\nu_e$&$\bar{\nu}_e$\\ \hline\hline
15.7&10.7 \\ \hline
1.5&0.6\\ \hline
1.0&0.4\\
\end{tabular}
&
\begin{tabular}{|c|c|}
\hline
$\nu_e$&$\bar{\nu}_e$\\ \hline\hline
14.8&10.7 \\ \hline
1.2&0.5\\ \hline
0.8&0.4\\
\end{tabular}
&
\begin{tabular}{|c|c|}
\hline
$\nu_e$&$\bar{\nu}_e$\\ \hline\hline
15.3&10.8 \\ \hline
1.0&1.0\\ \hline
0.7&0.7\\
\end{tabular}
&
\begin{tabular}{|c|c|}
\hline
$\nu_e$&$\bar{\nu}_e$\\ \hline\hline
15.3&10.3 \\ \hline
1.0&0.7\\ \hline
0.7&0.5\\
\end{tabular}
\\ \hline
\end{tabular}
\caption{Relic Number Fluxes (cm$^{-2}$s$^{-1}$) 
of electron neutrinos and anti-neutrinos, for different neutrino energy ranges and in the MSW case (no interaction with the magnetic field is included). The different cases correspond to normal (NH) or inverted (IH) hierarchy and 
sin$^2 2\theta_{13}=0.19$ (L) or  
sin$^2 2\theta_{13}=10^{-4}$ (S).
\label{tab:rflux} }
\end{center}
\end{table}

\subsection{Events in future observatories}
\noindent
We consider both the electron anti-neutrino signal associated to scattering on protons and the electron neutrino signal
due to neutrino capture on nuclei. We take oxygen, carbon and argon as nuclear targets, having in mind the MEMPHYS (UNO, Hyper-K), LENA and GLACIER detectors at present under study \cite{Autiero:2007zj}. 
The reactions of interest 
are\footnote{Note that we do not consider here events produced 
by the $\bar{\nu}_e + ^{12}$C$ \rightarrow e^{+} + ^{12}$B reaction
in LENA, since the cross section and the 
$\bar{\nu}_e$ flux are smaller than in the $\nu_e$ case. For GLACIER, 
we do not consider other channels since the charged-current  $\nu_e$ 
is the dominating one.}: $\bar{\nu}_e + p \rightarrow n + e^{+}$ 
and $\nu_e + ^{16}$O$ \rightarrow e^{-} + $X or
 $\nu_e + ^{12}$C$ \rightarrow e^{-} + ^{12}$N  for the 
 water and scintillator detectors respectively; 
$\nu_e + ^{40}$Ar$ \rightarrow e^{-} + ^{40}$K for liquid argon. While the
events associated to $^{16}$O cannot be 
distinguished from the ones on protons in water \v{C}erenkov detectors,
those corresponding to $^{12}$C 
can be identified by performing a double coincidence with the electrons
from $^{12}$N 
beta decay.
It is clear that the both $\bar{\nu}_e + p \rightarrow n + e^{+}$ and the $\nu_e + ^{40}$Ar cross sections have the advantage of having a very low threshold and of being much larger compared to nuclear targets such as 
$^{12}$C and $^{16}$O (Figure \ref{fig:cross}). 
It is however important to emphasize that while the neutrino-proton cross section is known with high accuracy, this is not the case for neutrino-nucleus interactions. 
One might hope that by the time large-scale observatories are in operation, new facilities producing neutrinos in the 100 MeV energy range
such as low energy beta-beams \cite{Volpe:2003fi} or those based on conventional sources \cite{efremenko} might provide us with precise neutrino-nucleus cross sections. 
\begin{figure}[t]
\vspace{.6cm}
\centerline{\includegraphics[scale=0.3,angle=-90]{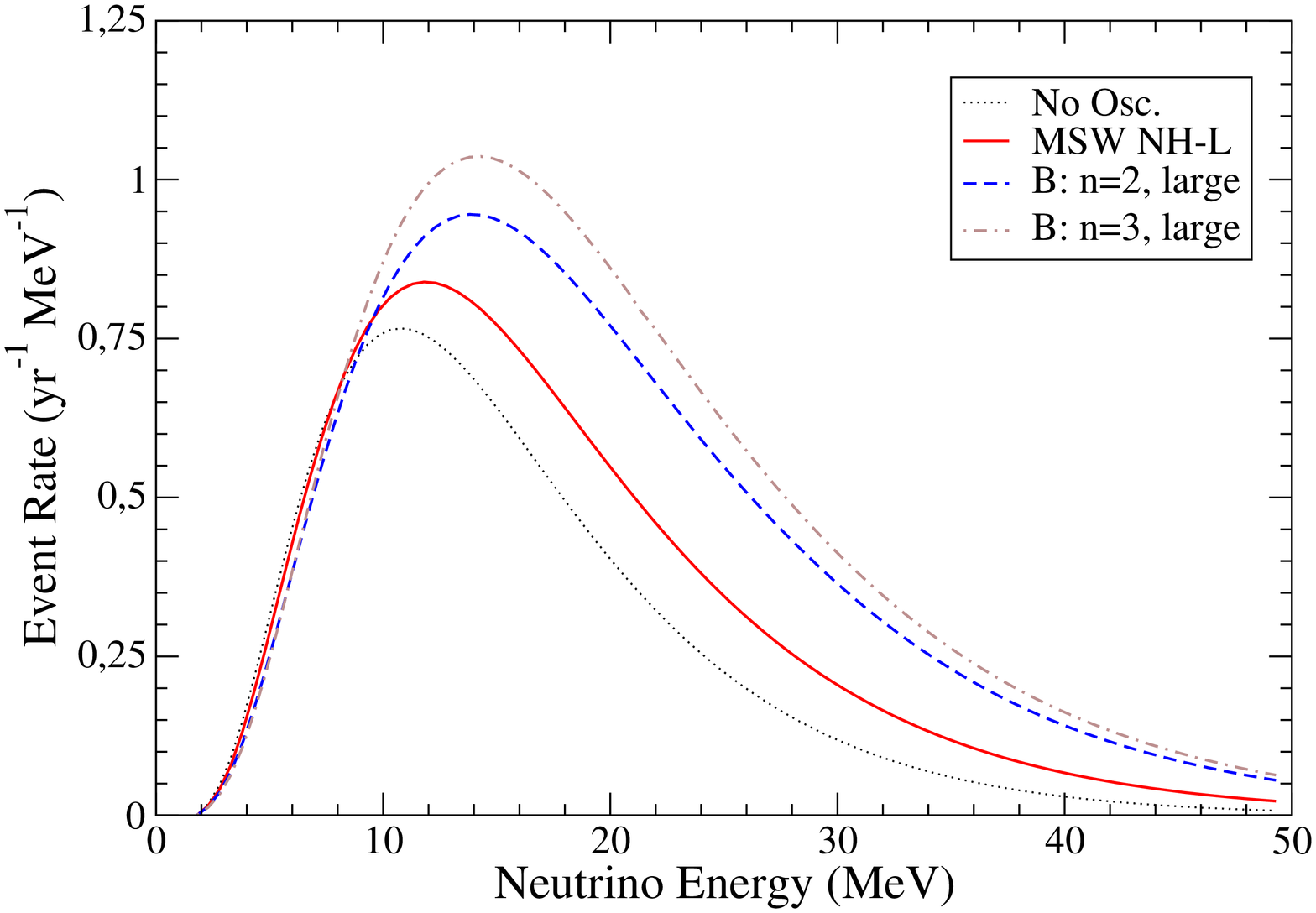}
\hspace{.2cm}\includegraphics[scale=0.3,angle=-90]{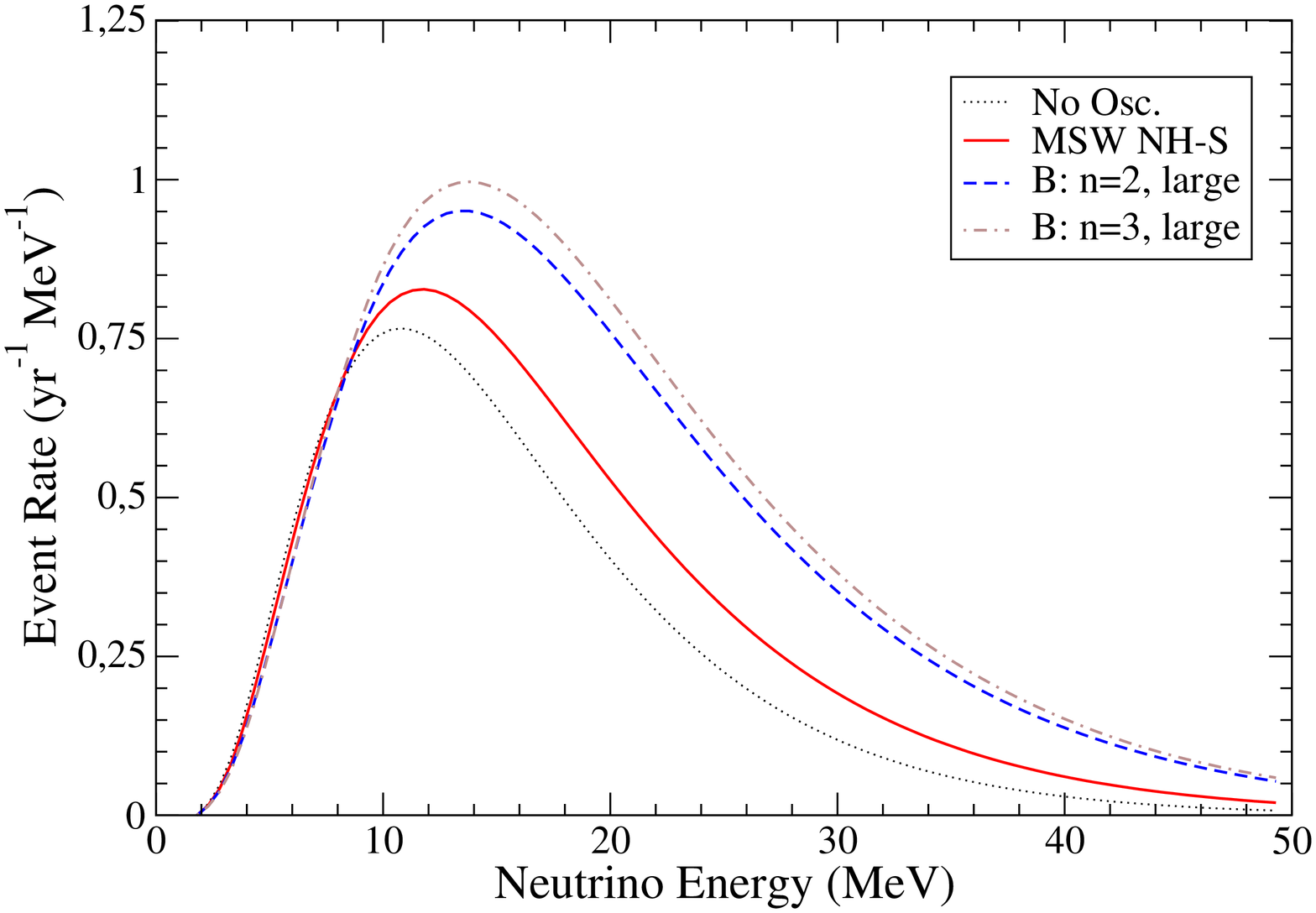}}
\caption{Event rates (100 kton $\times$ year)$^{-1}$ for $\bar{\nu}_e + p \rightarrow n + e^{+}$ due to SRN for  
a normal hierachy and a large (left) or small (right) value of $\theta_{13}$. The various curves present the results 
when the interaction with matter (MSW) and the coupling of the neutrino magnetic moment to the stellar magnetic fields are
included (B large) with a radial dependence of r$^{-2}$ (n=2) or r$^{-3}$ (n=3) Eq.(\ref{e:5}). The expected events when no oscillation is taken into account are shown for comparison. \label{fig:memphys-NH}}
\end{figure}

\begin{figure}[t]
\vspace{.6cm}
\centerline{\includegraphics[scale=0.3,angle=-90]{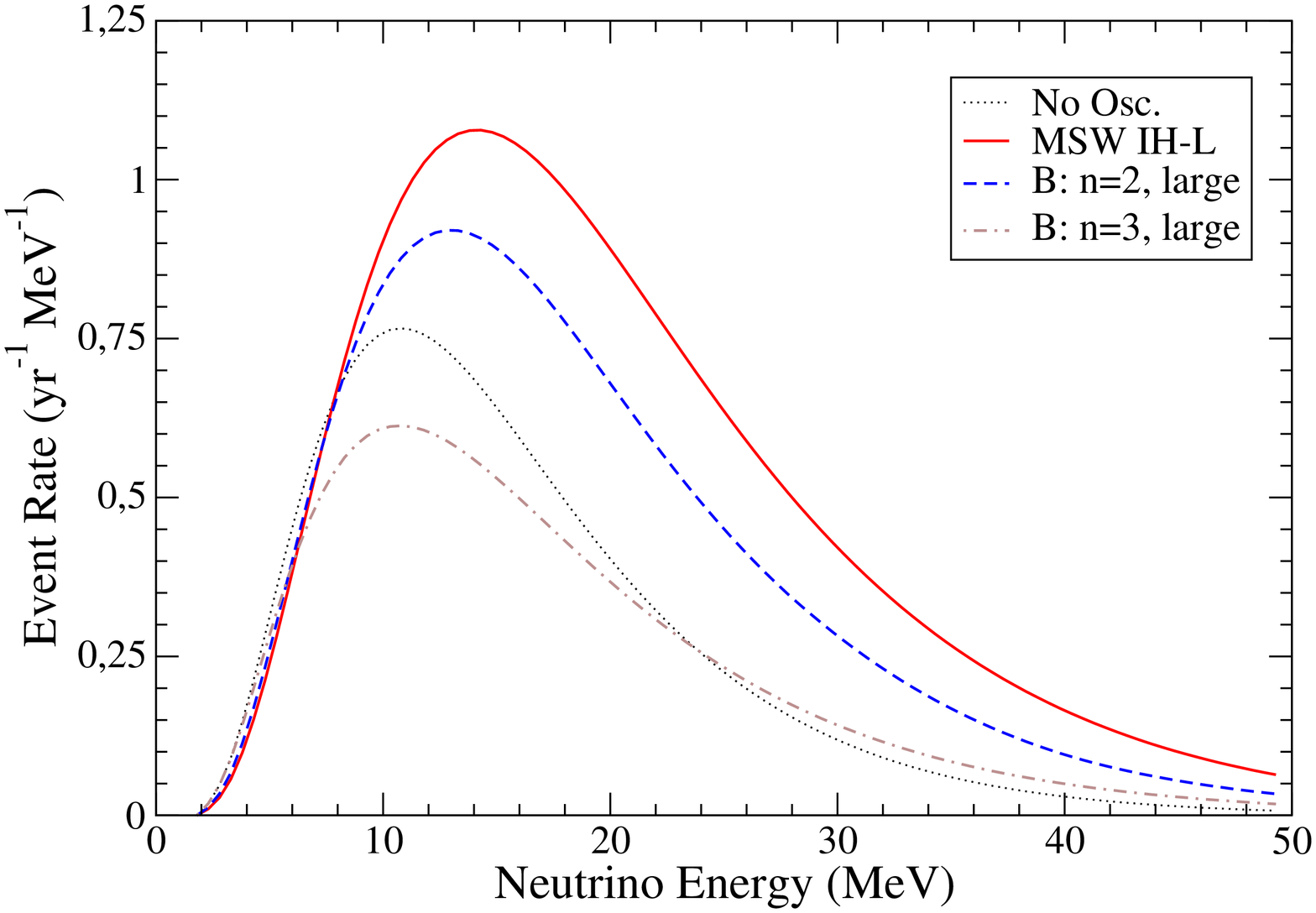}
\hspace{.2cm}\includegraphics[scale=0.3,angle=-90]{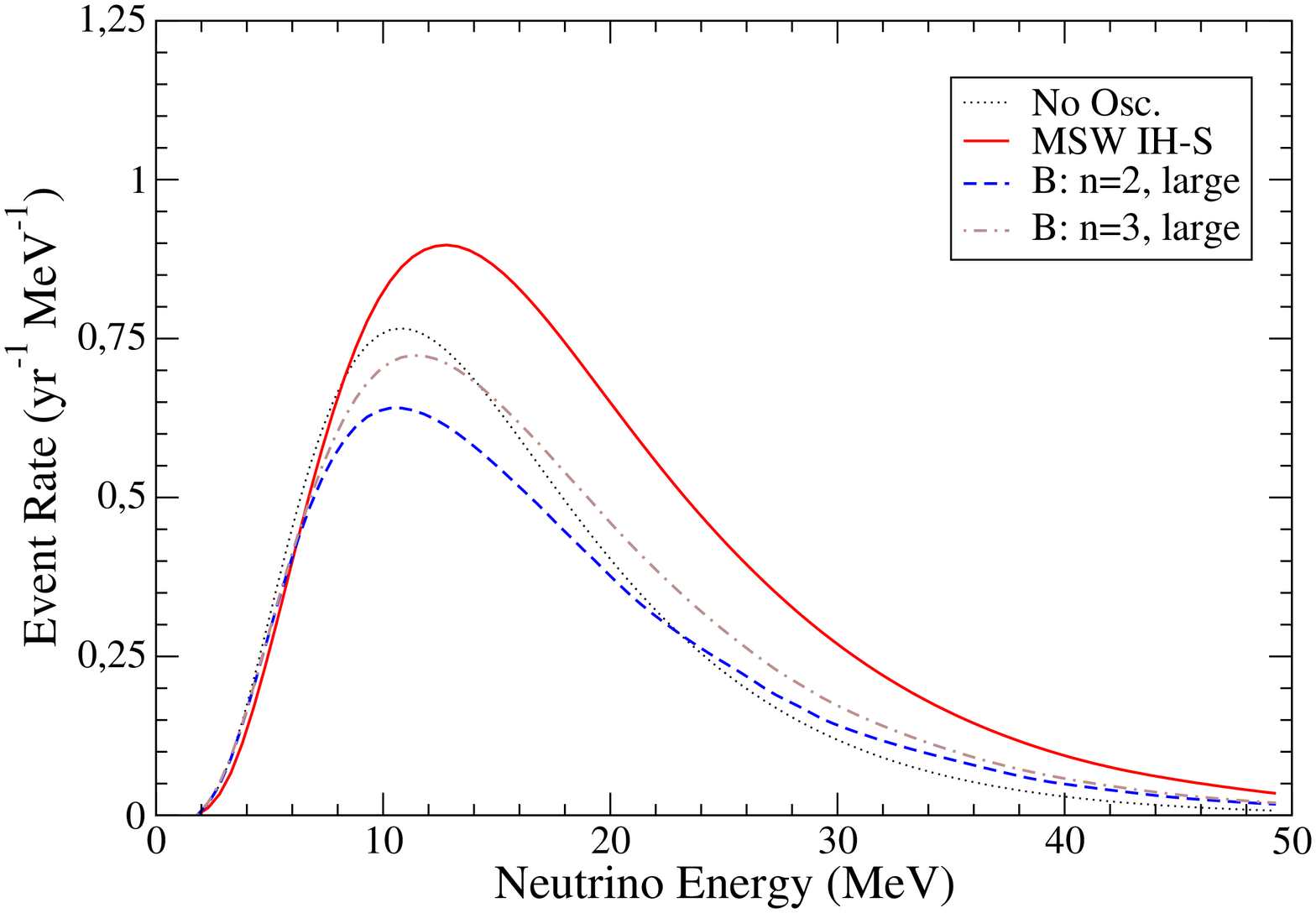}}
\caption{Same as Figure \ref{fig:memphys-NH} but for the inverted hierarchy.
\label{fig:memphys-IH}}
\end{figure}

\begin{figure}[t]
\vspace{.6cm}
\centerline{\includegraphics[scale=0.3,angle=-90]{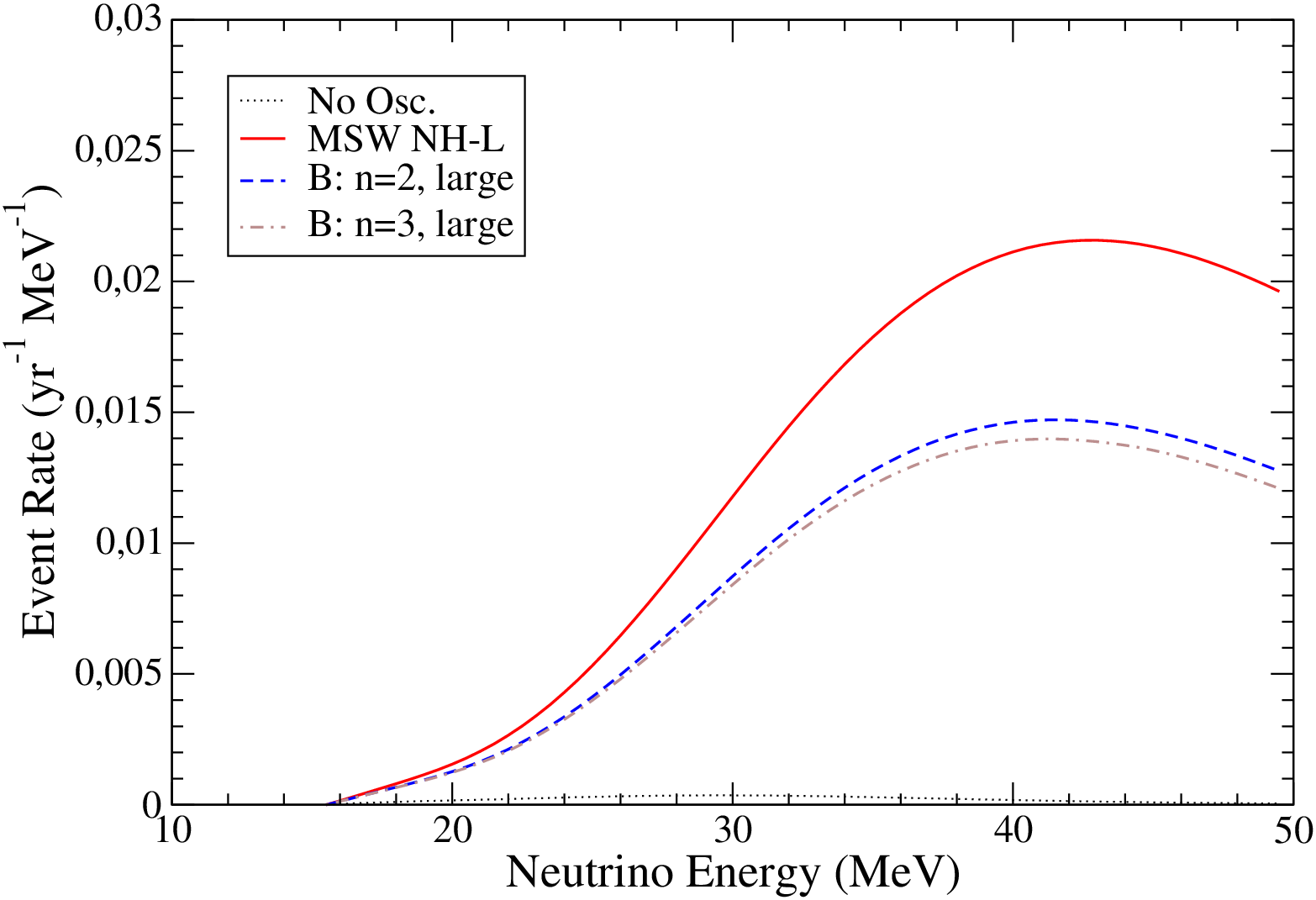}
\hspace{.2cm}\includegraphics[scale=0.3,angle=-90]{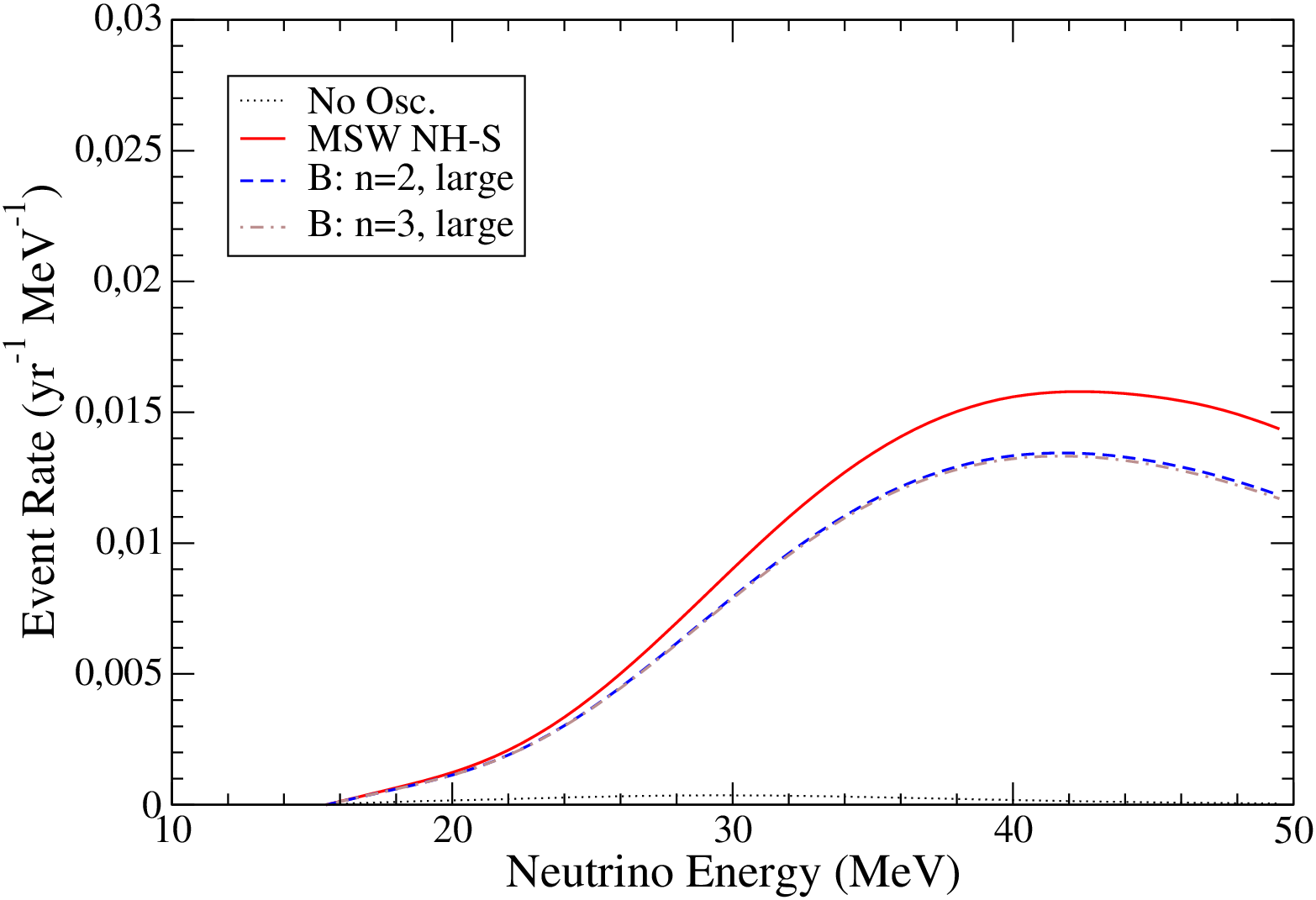}}
\caption{Same as Figure \ref{fig:memphys-NH} but for the event rates (100 kton $\times$ year)$^{-1}$ associated to 
$\nu_e + ^{16}$O$ \rightarrow e^{-} + X $. 
\label{fig:o-NH}}
\end{figure}

\begin{figure}[t]
\vspace{.6cm}
\centerline{\includegraphics[scale=0.3,angle=-90]{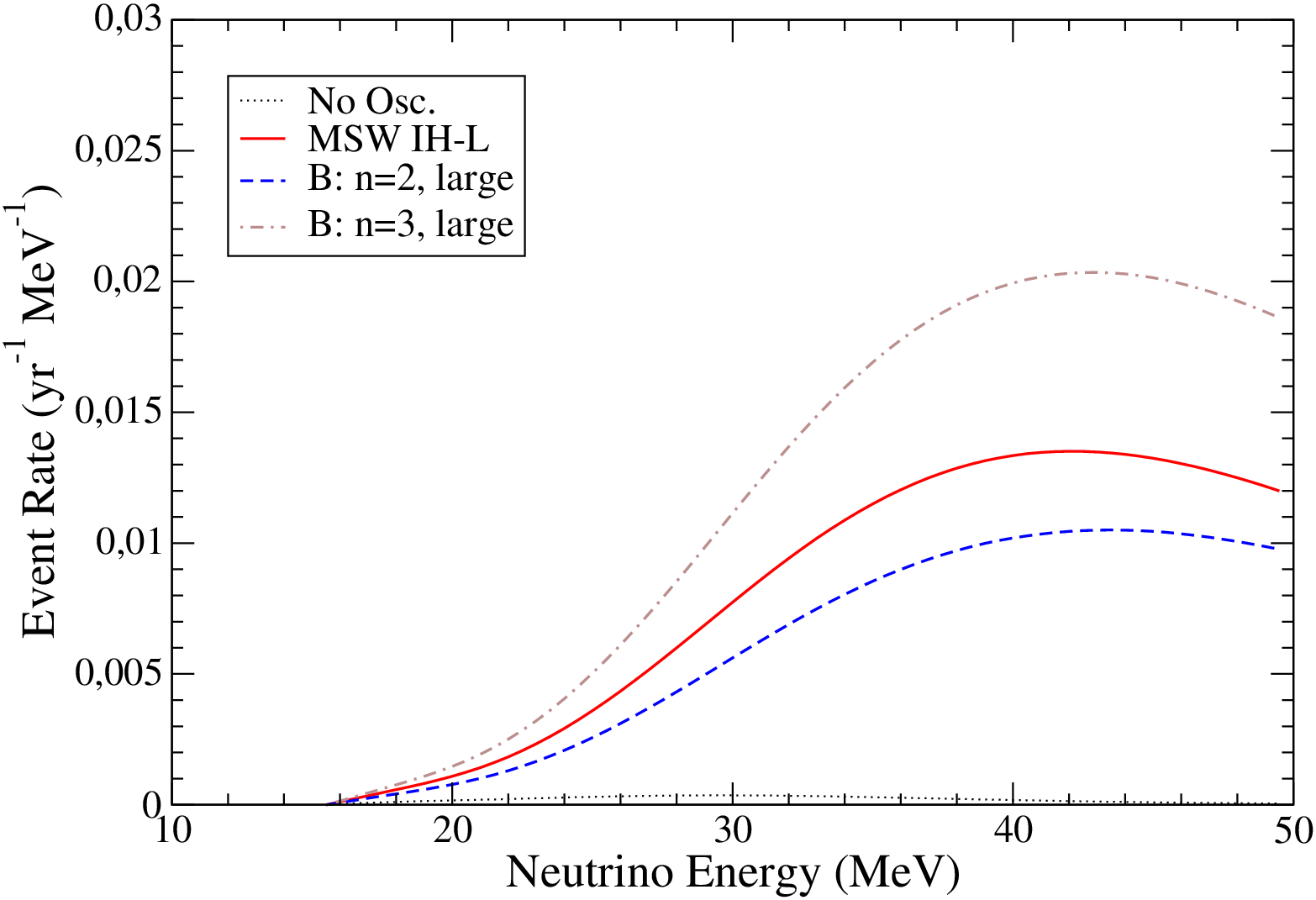}
\hspace{.2cm}\includegraphics[scale=0.3,angle=-90]{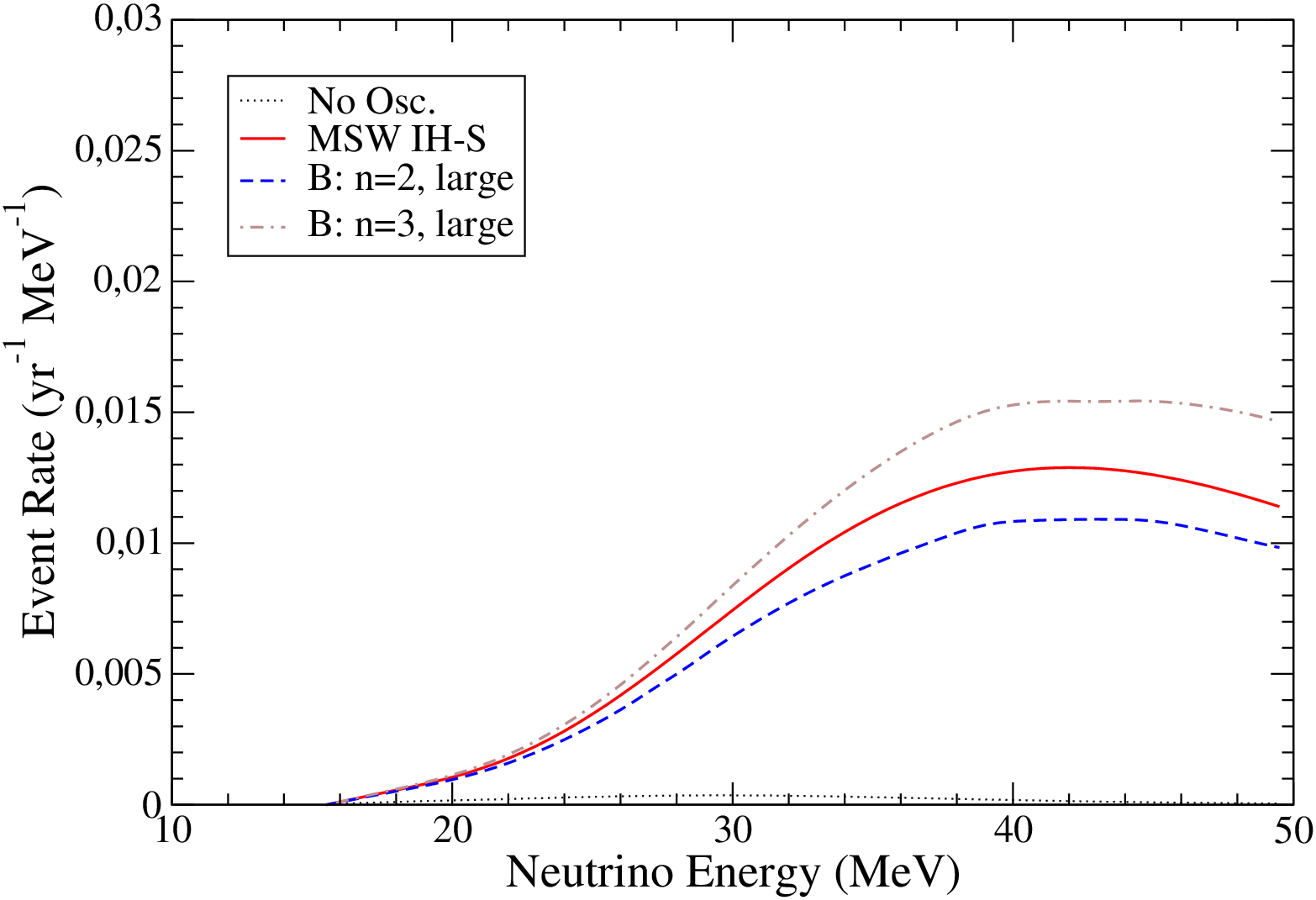}}
\caption{Same as Figure \ref{fig:o-NH} but for the inverted hierarchy.
\label{fig:o-IH}}
\end{figure}

\begin{figure}[t]
\vspace{.6cm}
\centerline{\includegraphics[scale=0.3,angle=-90]{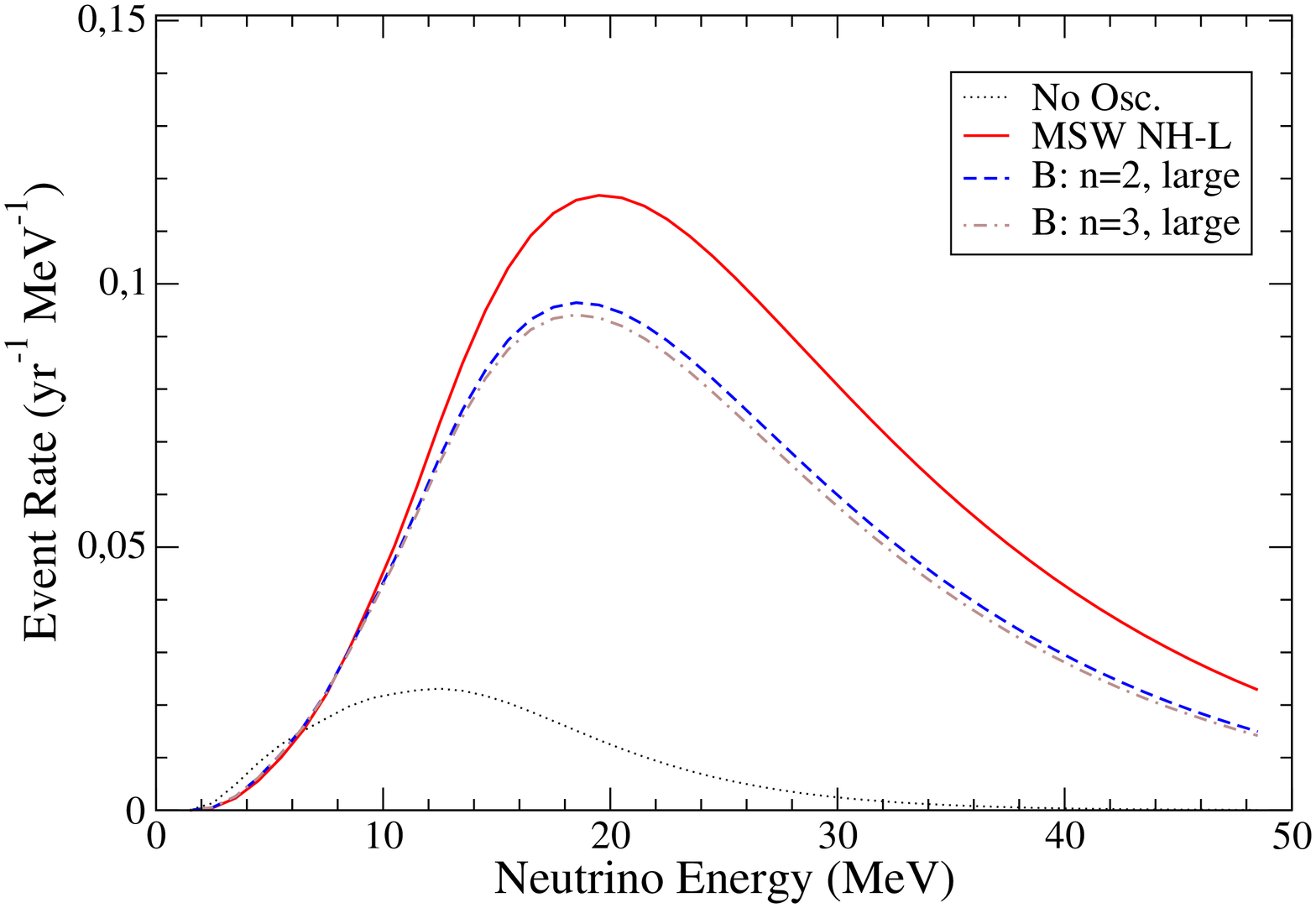}
\hspace{.2cm}\includegraphics[scale=0.3,angle=-90]{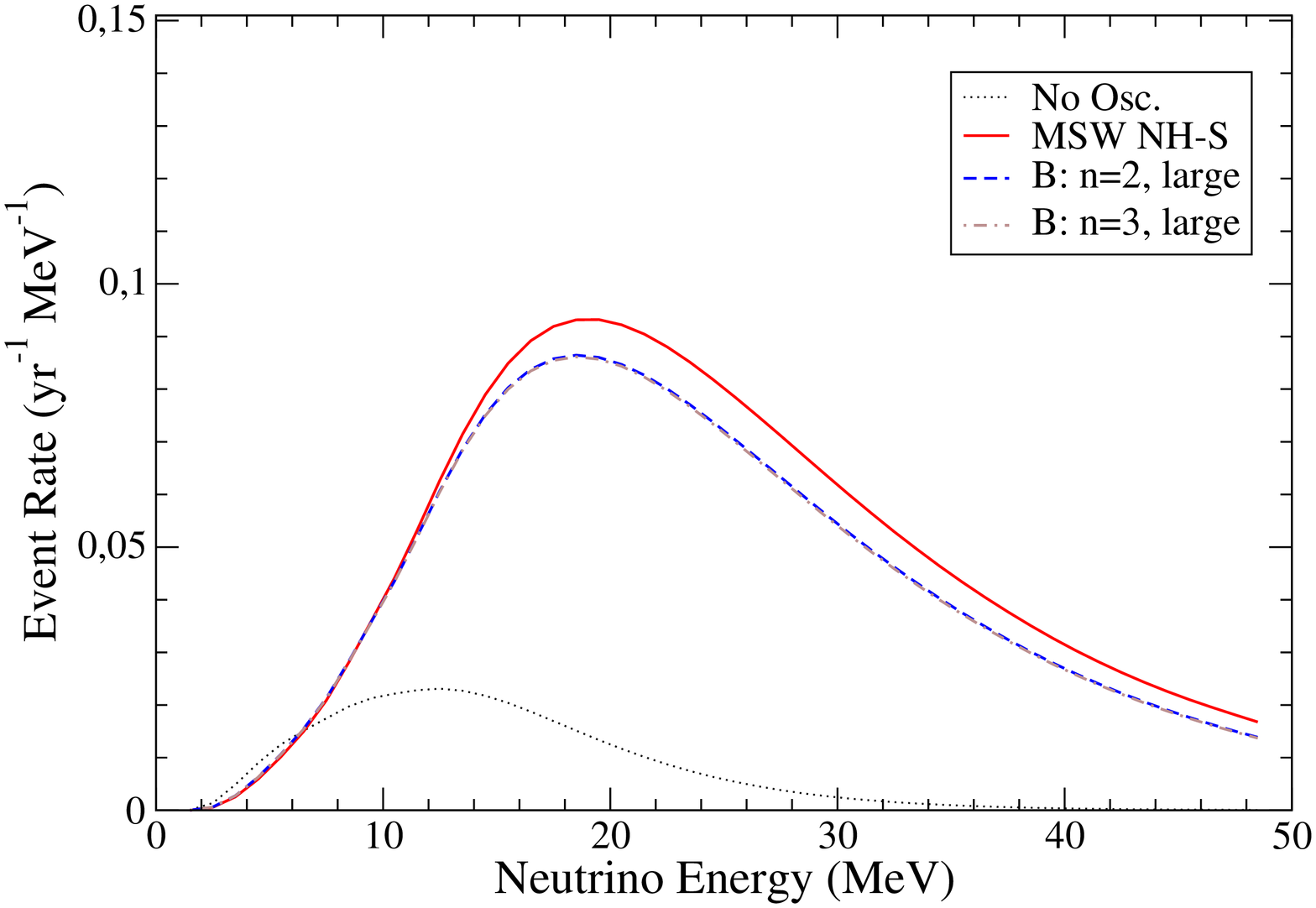}}
\caption{Same as \ref{fig:memphys-NH} but for the event rates (10 kton $\times$ year)$^{-1}$ associated to $\nu_e + ^{40}$Ar$ \rightarrow e^{-} + ^{40}$K \label{fig:ar-NH}.}
\end{figure}

\begin{figure}[t]
\vspace{.6cm}
\centerline{\includegraphics[scale=0.3,angle=-90]{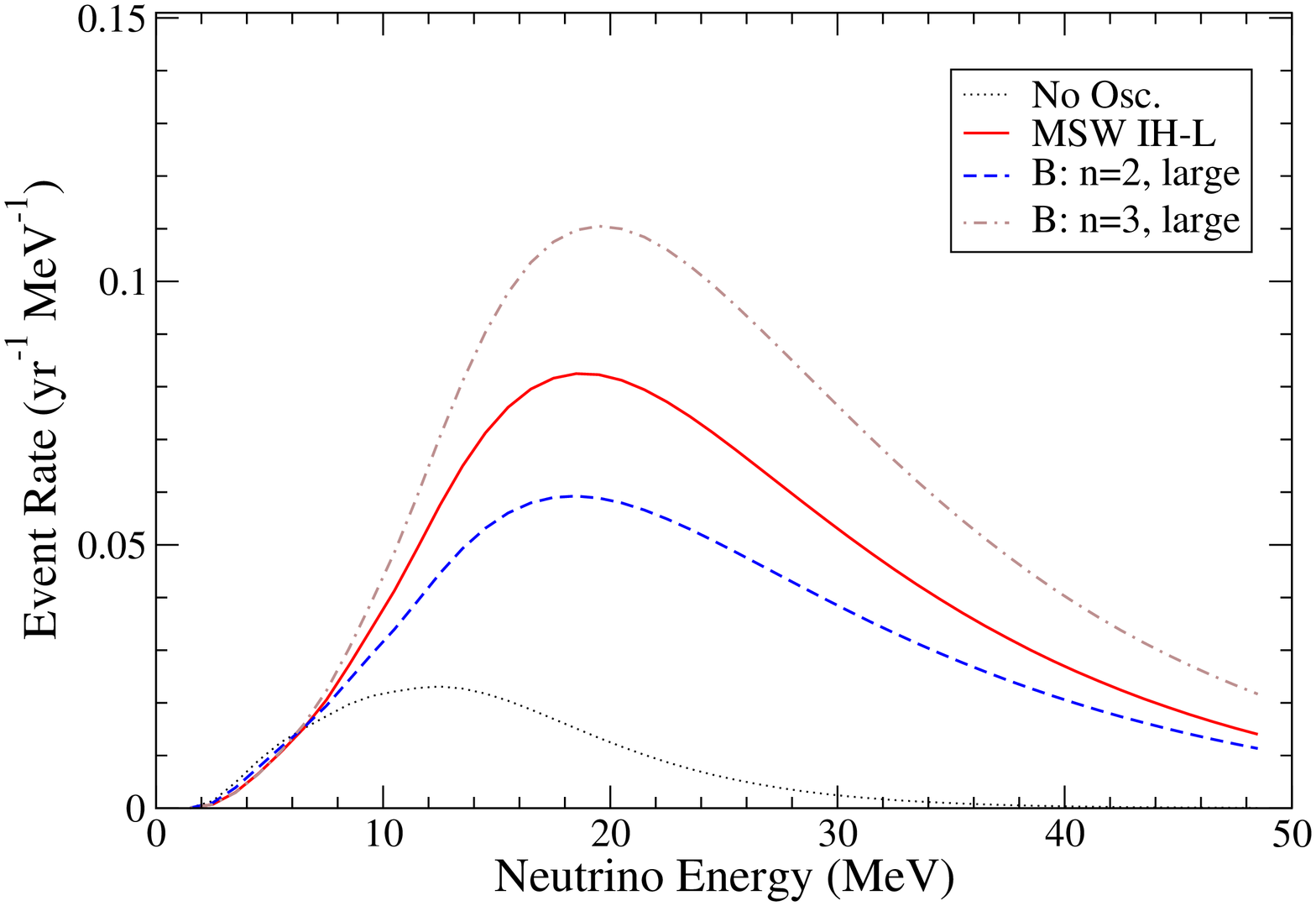}
\hspace{.2cm}\includegraphics[scale=0.3,angle=-90]{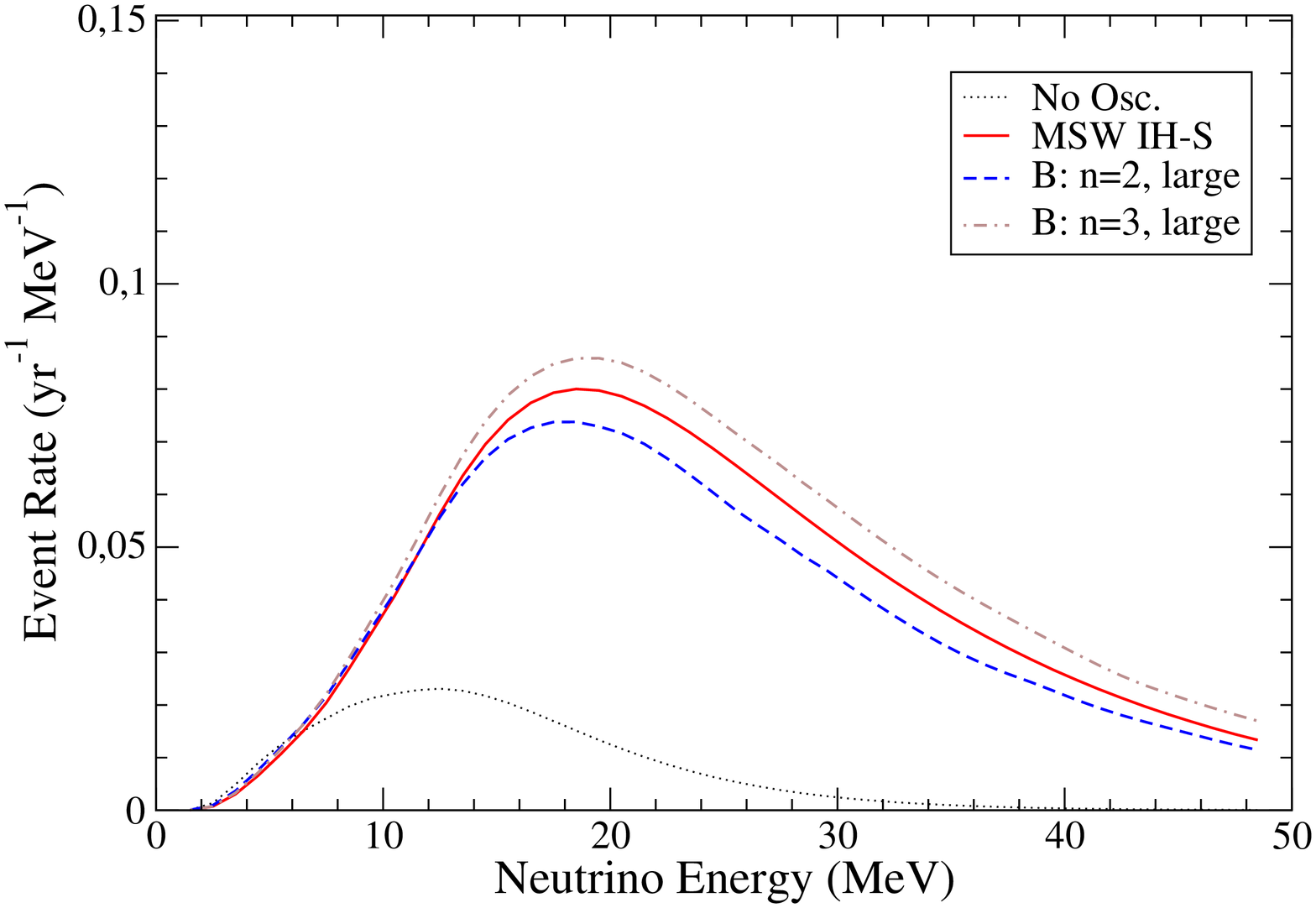}}
\caption{Same as Figure \ref{fig:ar-NH} but for the inverted hierarchy.
\label{fig:ar-IH}}
\end{figure}

Figures \ref{fig:memphys-NH}-\ref{fig:memphys-IH}, \ref{fig:o-NH}-\ref{fig:o-IH} and \ref{fig:ar-NH}-\ref{fig:ar-IH}  
present the event rates for scattering of $\bar{\nu}_e$ on protons and $\nu_e$ on oxygen and argon, respectively. 
These curves can be easily understood from the the relic number fluxes 
(Figs. \ref{fig:RSN-NH} to \ref{fig:RSNnuebar-IH}) and the cross sections (Figure \ref{fig:cross}),
since the neutrino-nucleon and neutrino-nucleus cross sections behave approximately like $E_{\nu}^2$.
The corresponding expected number of events for $\bar{\nu}_e$ and $\nu_e$ in 10 years are shown in 
Tables \ref{tab:p-NH}-\ref{tab:p-IH}\footnote{Note that the
$\bar{\nu}_e + $p
event rates presented here for LENA are
obtained with $f_{\star}=1$ while in \cite{Wurm:2007cy} with $f_{\star}=2.5$.}
and  \ref{tab:n-NH}-\ref{tab:n-IH}
for the water \v{C}erenkov (MEMPHYS, UNO, Hyper-K), scintillator (LENA\footnote{LENA is composed by 20 $\%$ PXE and 80 $\%$ Dodecane. The number of protons in the fiducial volume is 2.9  $\times$ 10$^{33}$.}) or liquid argon (GLACIER) technologies. The detector sizes are those considered in the corresponding proposals, namely 
440 kton fiducial mass (3 shafts), 50 kton and 100 kton respectively \cite{Autiero:2007zj}. 
Here we consider that each detector has 100 $\%$ efficiency.

\begin{figure}[t]
\vspace{.6cm}
\includegraphics[scale=0.3,angle=-90]{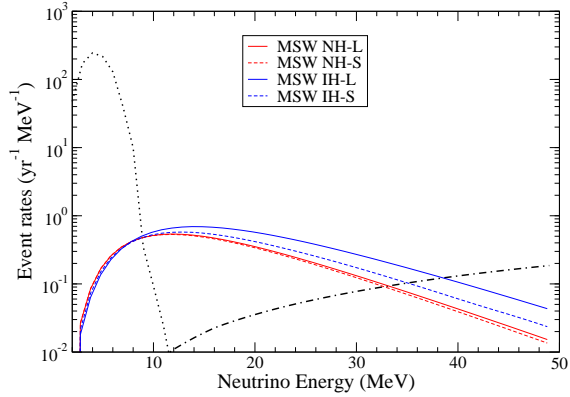}
\caption{Expected event rates in the LENA detector 
for normal (NH) and inverted  (IH) hierarchy as well as a small (S) and a large (L) 
third neutrino mixing angle. Both the reactor electron anti-neutrinos (dotted line) 
at Pyh\"asalmi and the atmospheric (dashed-dotted line) backgrounds are shown. 
The backgrounds are from Ref.\cite{Wurm:2007cy,Wurm}. \label{fig:backgrounds}}
\end{figure}

Extensive discussions about possible backgrounds can be found in the literature.
A detailed investigation is preformed in  \cite{Ando:2002ky} for water  \v{C}erenkov, in \cite{Wurm:2007cy} for scintillators and in \cite{Cocco:2004ac} for liquid argon.
This issue is clearly crucial for the observation of relic supernova neutrinos. 
Three are the main sources of the backgrounds: solar neutrinos, reactor anti-neutrinos, atmospheric neutrinos.
Solar neutrinos have low energies, the maximal being of almost 20 MeV coming from $hep$. Since water \v{C}erenkov detectors do not
have any charge identification, electron neutrinos can be mis-identified as anti-neutrinos. However, this background
can be almost fully suppressed thanks to the good angular resolution of such detectors. On the other hand,
the signal from electron neutrinos and anti-neutrinos can be distinguished in scintillator detectors. 
Concerning reactor electron anti-neutrinos, the flux is peaked at 2-3 MeV and has a tail reaching about 13 MeV. This contribution might be important or not depending on the chosen location.
If present, it can clearly hinder SRN coming from $ z\ge$1-2. 
In the case of the capture of the relic electron neutrinos by oxygen or carbon, the reaction threshold is so high that
neither reactor nor solar neutrinos can produce a background.
Atmospheric neutrinos produce events at higher
energy. Note that in water \v{C}erenkov detectors there is a supplementary source of background due to invisible muons. These are produced by atmospheric neutrinos that interact with the nucleons in the detector producing muons
with energies below the \v{C}erenkov radiation threshold. Figure 
\ref{fig:backgrounds} shows as an example the backgrounds in a scintillator detector such as LENA in comparison with our $\bar{\nu}_e$ rates.

Various authors have identified 
an energy window for the observation of relic supernova neutrinos, where the backgrounds are
lower than the expected signal. At the present stage it might be premature
to define an energy window for two reasons. First some backgrounds like for example the atmospheric one might scale differently in large-scale observatories from what is currently assumed. Secondly there are still many theoretical uncertainties coming either from supernova modelling, or from the knowledge of the supernova rate and from still unknown neutrino properties, in particular the knowledge of the hierarchy or of the
$\theta_{13}$ value. 
For the above mentioned reasons we present the expected rates in two different energy windows. 
The 6.8 and 5-50 MeV energy ranges\footnote{The threshold of 6.8 MeV takes
into account the 1.8 MeV threshold energy for the reaction
$\bar{\nu}_e +$p $\rightarrow $n + e$^{+}$ plus the 5 MeV necessary to
observe the outgoing electron in water \v{C}erenkov.
The threshold of 5 MeV for scintillators is fixed in the same way.}are taken to 
show how the event rate might increase if one manages to have a threshold
as low as 6.8 and 5 MeV for the outgoing lepton and as high as 50 MeV, by suppressing the atmospheric background.
Such an information, that might be helpful in the future, should be taken as indicative only.
The 19.3-27, 9.3-25, 17.5-41.5 MeV for water \v{C}erenkov, scintillators, liquid argon detectors are taken following Refs.
\cite{Ando:2002ky},\cite{Wurm:2007cy},\cite{Cocco:2004ac} respectively.  We make this choice to facilitate the
comparison with available predictions\footnote{Note that our relic fluxes cross the backgrounds at higher energy (see for example \ref{fig:backgrounds}). This would give energy ranges larger than the presented ones.}.

As can be seen from Tables \ref{tab:p-NH}-\ref{tab:p-IH}, 
in absence of the coupling to magnetic fields the largest number of events is expected for the IH-L case because of the MSW-$h$ resonance. The events are larger by 50-60 $\%$.
In the normal hierarchy ($\theta_{13}$ large or small) case,
electron anti-neutrinos undergoes regular matter effects and the RSF-$h$ resonance 
which renders the electron
anti-neutrinos hotter compared to the case with matters effects only. Therefore
in presence of strong magnetic fields and/or large values of $\mu_{ij}$ the event rates 
can be increased by 30 to 50 $\%$.
For the inverted hierarchy, electron anti-neutrinos encounter 
both the MSW-$h$ and the RSF-$h$ resonances. In this case the events are significantly reduced. 
The two magnetic field profiles considered give similar results, except for the inverted hierarchy and large mixing angle.
For this case and if the $n=2$ profile is taken, the anti-neutrinos undergo the RSF-$l$ resonance as well. As a consequence
the number of events is doubled compared to the $n=3$ profile and is close to the MSW result.

In Tables \ref{tab:n-NH}-\ref{tab:n-IH} we show the predicted events in the case of electron neutrino-nucleus scattering, choosing
large intervals only\footnote{The 20.5 MeV threshold includes the reaction threshold and 5 (3) MeV electron energy for MEMPHYS (LENA).}, except for the argon case. In fact, this is the first time that interactions of SRN on oxygen and carbon are considered. Since a detailed study of the atmospheric backgrounds for these cases does not exist, we hope that our work will trigger investigations of this channels for MEMPHYS (UNO, Hyper-K) and LENA.
A few events are found for $\nu_e + ^{12}$C, up to 20 
for $\nu_e + ^{16}$O, and 100-200 for 
$\nu_e + ^{40}$Ar.  
Note that the large number of events found for $\nu_e + ^{40}$Ar is due both to the large cross section (Figure \ref{fig:cross}) and to the large $\nu_e$ flux (Table \ref{tab:rflux}).
In absence of magnetic moment effects, 
the maximal conversion for electron neutrinos occurs for a normal hierarchy and a large $\theta_{13}$ value, 
due to the MSW-$h$ resonance. In fact,
since $\nu_e$ gets a hotter spectrum because of the mixing with $\nu_{\mu},\nu_{\tau}$ neutrinos the number of events is 
increased by 30-60 $\%$ compared to the case when the resonance is not (fully) adiabatic or of normal matter effects.
On the other hand, RSF conversions from $\nu_e$ to $\bar{\nu}_{\mu},\bar{\nu}_{\tau}$ generally decrease by 20-40 $\%$, compared to the MSW only, in most cases. The only exception is for the inverted hierarchy and the $n=3$ magnetic field profile for which the events increase by 15 up to 50 $\%$.  It is clear that the search for relic $\nu_e$ might either improve the 
present LSD limit \cite{Aglietta:1992yk} (in case of non-observation) or bring crucial information to our understanding of core-collapse supernova and neutrino physics.

\begin{center}
\begin{table}[h!]
\begin{tabular}{|ccc|}
\hline
&MEMPHYS &LENA \\
&$\bar{\nu}_e$+p&$\bar{\nu}_e$+p \\
\begin{tabular}{|c|}
\hline
\\ \hline\hline
MSW NH-L\\ 
$(\mu B)_{large}$ NH-L\\  \hline\hline
MSW NH-S\\ 
$(\mu B)_{large}$ NH-S\\
\end{tabular}
&
\begin{tabular}{|c|c|}
\hline
6.8 - 50$\ $&19.3 - 27$\ $\\ \hline\hline
652 & 142 \\ 
873 (968) &214 (240)\\ \hline\hline
631&135\\ 
868 (921) &209 (225) \\ 
\end{tabular}
&
\begin{tabular}{|c|c|}
\hline
5.0 - 50 $\ $ &9.3 - 25$\ $  \\ \hline\hline
67 &44 \\ 
89 (97) &55 (61) \\\hline\hline
65 &43 \\
88 (94) &55 (58) \\
\end{tabular}
\\ \hline
\end{tabular}
\caption{Expected number of events after 10 years for electron 
anti-neutrino scattering on protons in MEMPHYS (water \v{C}erenkov) 
or LENA
(scintillator) detectors. 
The size considered is 440 kton fiducial mass, 50 kton and 100 kton respectively, 
following the corresponding proposals \cite{Autiero:2007zj}. 
An efficiency of 100 $\%$ is assumed.
The calculations correspond to the normal hierarchy (NH) and large (L) or small (S) value of the third neutrino mixing angle. The calculations include the coupling of 
neutrinos to matter effects without (MSW) or with the coupling 
to the stellar magnetic fields ($(\mu B)_{large}$) . The stellar magnetic 
profile is taken to be a power-law Eq.(\ref{e:5})
with $n=2$ ($n=3$ results 
are given in parenthesis). Two energy intervals are considered (see text).
\label{tab:p-NH}}
\end{table}
\end{center}

\begin{center}
\begin{table}[h!]
\begin{tabular}{|ccc|}
\hline
&MEMPHYS &LENA\\
&$\bar{\nu}_e$+p&$\bar{\nu}_e$+p \\
\begin{tabular}{|c|}
\hline
\\ \hline\hline
MSW IH-L\\ 
$(\mu B)_{large}$ IH-L\\ \hline\hline
MSW IH-S\\ 
$(\mu B)_{large}$ IH-S\\ 
\end{tabular}
&
\begin{tabular}{|c|c|}
\hline
6.8 - 50$\ $&19.3 - 27$\ $\\ \hline\hline
997&248\\ 
776 (464) &181 (95) \\ \hline\hline
751&173 \\ 
479 (559) &98 (120)\\ 
\end{tabular}
&
\begin{tabular}{|c|c|}
\hline
 5.0-50$\ $ &9.3-25$\ $  \\ \hline\hline
101 &64\\ 
80 (48)&51 (30)\\ \hline\hline
77& 50\\ 
51 (58) &32 (38)\\
\end{tabular}
\\ \hline
\end{tabular}
\caption{Same as Table \ref{tab:p-NH} but for the inverted hierarchy. \label{tab:p-IH}}
\end{table}
\end{center}

\begin{center}
\begin{table}[h!]
\begin{tabular}{|cccc|}
\hline
&MEMPHYS&LENA&GLACIER \\
&$\nu_e+\ ^{16}$O&$\nu_e+\ ^{12}$C&$\nu_e+\ ^{40}$Ar \\
\begin{tabular}{|c|}
\hline
Energy Range (MeV)$\ $\\ \hline\hline
MSW NH-L\\
$(\mu B)_{large}$ NH-L\\ \hline\hline
MSW NH-S\\ 
$(\mu B)_{large}$ NH-S\\
\end{tabular}
&
\begin{tabular}{|c|}
\hline
20.5 - 50$\ $ \\ \hline\hline
19\\
14 (13)\\\hline\hline
14\\
12 (12)\\
\end{tabular}
&
\begin{tabular}{|c|}
\hline
20.5 - 50$\ $ \\ \hline\hline
2.9\\
2.1 (2.0)\\\hline\hline
2.2\\
2.0 (1.9) \\
\end{tabular}
&
\begin{tabular}{|c|c|}
\hline
4.5 - 50$\ $ &17.5-41.5\\ \hline\hline
297&197\\
235 (228) &141(156)\\\hline\hline
235&152\\
213 (212)&136(135)\\
\end{tabular}
\\ \hline
\end{tabular}
\caption{Expected number of events after 10 years for electron neutrino scattering on nuclei in MEMPHYS (water \v{C}erenkov), LENA (scintillator) or GLACIER (liquid argon) detectors. The size considered is 440 kton, 50 kton and 100 kton respectively, following the corresponding proposals \cite{Autiero:2007zj}. An efficiency of 100 $\%$ is assumed.
The calculations correspond to the normal hierarchy (NH) and large (L) or small (S) value of the third neutrino mixing angle. The calculations include the coupling of neutrinos to matter effects without (MSW) or with the coupling 
to the stellar magnetic fields ($(\mu B)_{large}$) . The stellar magnetic profile is taken to be a power-law Eq.(\ref{e:5})
with $n=2$ ($n=3$ results are given in parenthesis). Two energy intervals are considered in the case of argon (see text).\label{tab:n-NH}}
\end{table}
\end{center}

\begin{center}
\begin{table}[h!]
\begin{tabular}{|cccc|}
\hline
&MEMPHYS&LENA &GLACIER\\
&$\nu_e+\ ^{16}$O&$\nu_e+\ ^{12}$C&$\nu_e+\ ^{40}$Ar \\
\begin{tabular}{|c|}
\hline
Energy Range (MeV)$\ $\\ \hline\hline
MSW IH-L\\ 
$(\mu B)_{large}$ IH-L\\ \hline\hline
MSW IH-S\\ 
$(\mu B)_{large}$ IH-S\\ 
\end{tabular}
&
\begin{tabular}{|c|}
\hline
$\ $20.5 - 50$\ $ \\ \hline\hline
12\\
9 (18)\\\hline\hline
12\\
10 (14)\\
\end{tabular}
&
\begin{tabular}{|c|}
\hline
$\ $20.5 - 50$\ $ \\ \hline\hline
1.9\\
1.4 (2.8)\\\hline\hline
1.8\\
1.5 (2.1) \\
\end{tabular}
&
\begin{tabular}{|c|c|}
\hline
$\ $4.5 - 50$\ $&17.5-41.5\\ \hline\hline
207&132\\
155 (282)&95 (186) \\\hline\hline
200&127\\
181 (221)&112(142) \\
\end{tabular}
\\  \hline
\end{tabular}
\caption{Same as Table \ref{tab:n-NH} but for the inverted hierarchy. \label{tab:n-IH}}
\end{table}
\end{center}

\section{Conclusions}
\noindent
In this work we have analysed both the relic supernova electron anti-neutrino and neutrino fluxes and the associated signals due to scattering on protons and nuclei respectively, in large-scale observatories such as MEMPHYS (UNO, Hyper-K), LENA or GLACIER. 
Our calculations of the neutrino propagation in the star include both the
matter interaction and the resonant spin-flavour conversion. One of our goals has been to try to investigate under which conditions
(profile, $\mu_{ij}$ and/or $B_0$ values) RSF effects could be seen.
In all cases studied, the results show that RSF phenomenon might increase/decrease significantly the number of events, for strong magnetic fields and/or large values of the neutrino magnetic moment. It is interesting to note that these effects are maximal for electron anti-neutrinos, when no resonant conversion is present due to matter efffects (normal hierarchy and a large or small value of $\theta_{13}$) due to the RSF-$h$ resonance. In the case of electron neutrinos, the RSF tend to decrease the events, in most cases. Our results are also sensitive to the magnetic field profiles. 
To obtain thes results with the spin-flavour conversion, we have made the assumption of a static density supernova profile and also neglected possible effects of the shock wave on the stellar magnetic fields.
Our present knowledge in this respect is in fact still 
too limited to include such aspects. From this point of view, 
our analysis should be taken as a first step in this direction and should be 
updated once progress is performed on the knowledge of the supernova
magnetic fields and on the shock wave impact.  

Our main goal has been to perform a numerical study of the relic
fluxes and their signal in the next generation observatories.
We have obtained that $\nu_e$ fluxes are larger than the $\bar{\nu}_e$ up to a factor 2.5 depending on the oscillation scenarios (normal or inverted hierarchy and $\theta_{13}$ value). We have shown that,   
after 10 years operation, between 100 and 250 $\bar{\nu}_e + p$ events are expected in the 19.3-27 MeV neutrino energy range for MEMPHYS, while 45-65 are expected in LENA in the 9.3-25 MeV energy range. We find that neutrino-carbon scattering will produce 2-3 events, while neutrino-oxygen will give 10-20 events and the neutrino-argon interaction will induce 100 to 200 events.
(These number are obtained by taking the normalization factor for the star formation rate 
$f_{\star}=1$, which implies a local formation rate of 0.7$\times 10^{-2} h_{70} M_{\odot}$ year$^{-1}$Mpc$^{-3}$, consistent with 
mildly dust-corrected UV data at low redshift).
The possible measurement of both $\nu_e$ and $\bar{\nu}_e$ could help us determining the neutrino fluxes at the neutrinosphere and/or still unknown neutrino properties, such as the hierarchy or the knowledge of the third neutrino mixing angle. This will 
depend on the progress made in supernova modelling or neutrino physics by the time such observatories might come into operation.
 In conclusion,
future large-scale observatories based on water, scintillator or liquid argon have a great discovery potential for relic supernova neutrinos. Their observation would clearly provide us with essential information on core-collapse supernova explosions, the supernova explosion rate and important neutrino properties that still remain unknown.

\section*{ACKNOWLEDGMENTS}
We are grateful to Takashi Yoshida for illuminating discussions and to Shin'ichiro Ando, Gail McLaughlin, Jim Kneller, Karlheinz Langanke, Teresa Marrodan Undagoitia, Mauro Mezzetto, Alessandra Tonazzo, 
Petr Vogel, Michael Wurm for providing us with useful information.
We acknowledge support from the project ANR-05-JC-JC0023 ``Non-standard neutrino
properties and their impact in astrophysics and cosmology'' (NeuPAC).

\end{document}